\documentclass[%
 reprint,
showpacs,
 amsmath,
 amssymb,
 aps,
prb,
nobalancelastpage]{revtex4-1}

\usepackage{color}
\usepackage{graphicx}
\usepackage{dcolumn}
\usepackage{bm}
\usepackage{esint}
\usepackage{todo}
\usepackage{hyperref}
\usepackage{cleveref}
\usepackage{csquotes}
\usepackage{verbatim}
\usepackage{hhline}
\usepackage[normalem]{ulem}

{\renewcommand{\arraystretch}{1.5}
}

\crefname{appsec}{appendix}{appendices}

\usepackage{soul}
\usepackage{cancel}

\begin{document}

\title{Thin film modeling of crystal dissolution and growth in confinement}
\author{Luca Gagliardi}
\email{luca.gagliardi@univ-lyon1.fr}
\author{Olivier Pierre-Louis}
\email{olivier.pierre-louis@univ-lyon1.fr}
\affiliation{CNRS, ILM Institut Lumi\`ere Mati\`ere,\\
Universit\' e Claude Bernard Lyon 1
Campus LyonTech-La Doua
Batiment Brillouin, 10 rue Ada Byron 
F-69622 Villeurbanne, France}
\date{\today}
\begin{abstract}

We present a continuum model describing dissolution
and growth of a crystal contact confined against a substrate. 
Diffusion and hydrodynamics in the liquid film separating the crystal 
and the substrate are modeled within the lubrication approximation. 
The model also accounts for the disjoining pressure and surface tension.
Within this framework, we obtain evolution equations 
which govern the non-equilibrium dynamics of the crystal interface.
Based on this model, we explore the problem 
of dissolution under an external load, known as pressure solution.
We find that in steady-state, diverging (power-law) crystal-surface 
repulsions lead to flat contacts with a monotonic 
increase of the dissolution rate as a function of the load.
Forces induced by viscous dissipation 
then surpass those due to disjoining pressure
at large enough loads.
In contrast,  finite repulsions (exponential) lead to sharp pointy contacts 
with a dissolution rate independent on the load and on the liquid viscosity. 
Ultimately, in steady-state the crystal never touches the substrate when pressed against it, 
independently from the nature of the crystal-surface interaction 
due to the combined effects of viscosity and surface tension.

\end{abstract}

\maketitle

\section{Introduction}\label{sec:intro}

\begin{figure}[b]
\includegraphics[width=\linewidth]{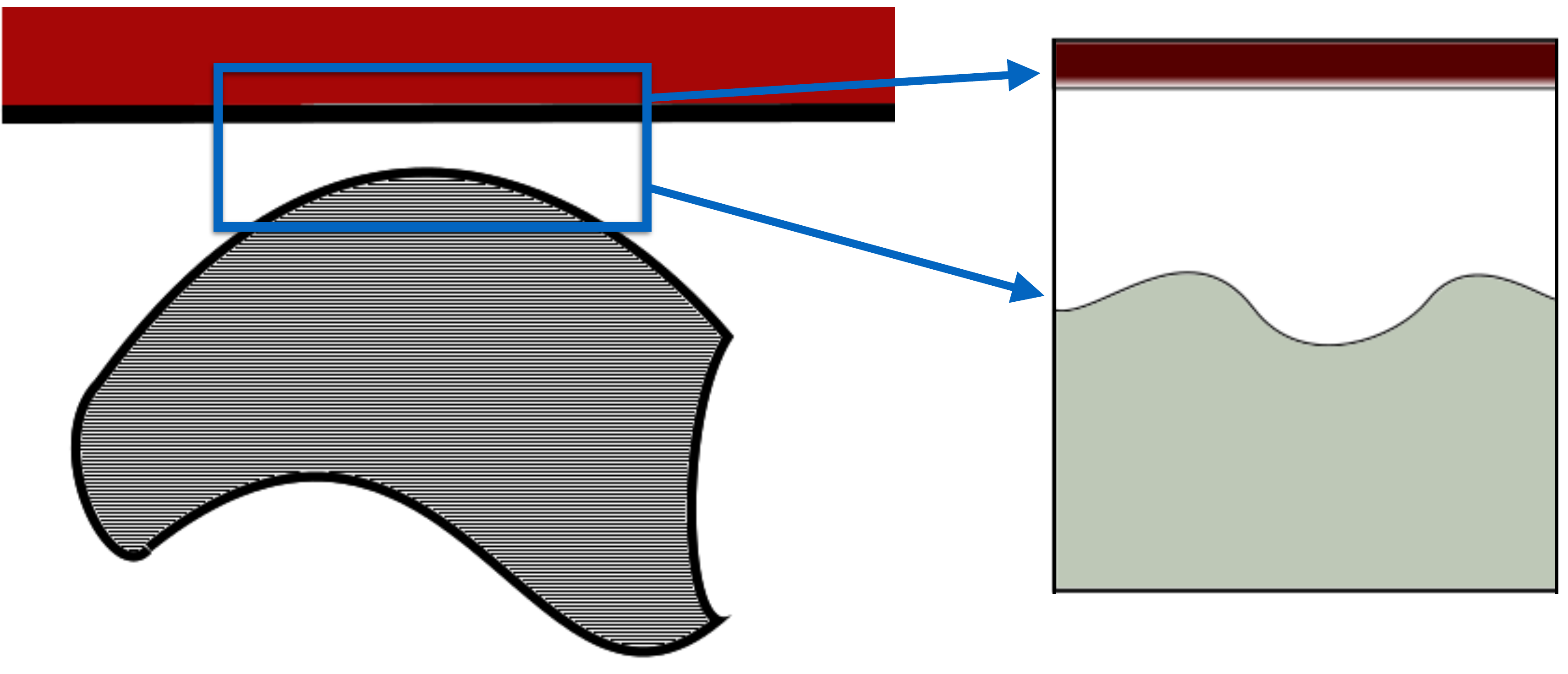}
\caption{Sketch of an arbitrary shaped crystal in the vicinity of a flat substrate. 
The panel on the right side presents a zoom of the contact region.}
\label{fig:model_sketch}
\end{figure}

Pressure solution is the stress-induced dissolution 
of solids in the presence of an applied load. This
phenomenon for example contributes to the evolution 
of the structure of sedimentary rocks 
which are initially constituted of porous
or loose  assemblies of grains.
When these rocks are under pressure, 
grains dissolve in areas of relatively high stresses 
located in contact regions between grains, 
and re-precipitate elsewhere, 
leading to a slow compaction of the global structure.
Hence, pressure solution is recognized as one of the main 
mechanism of rock diagenesis and metamorphism.
Pressure solution, and the opposite problem of crystallization force 
(the mechanical stress generated by crystal growth), 
 have attracted attention since the beginning of the 20th 
century~\cite{Lavalle1853,Sorby1856,Becker1905,Taber1916,Correns1939,Ramsay1968,Durney1972,Desarnaud2016a}.

The material dissolved during pressure solution
is usually evacuated away from the contact 
region towards a macroscopic reservoir
through the thin liquid film
between the crystal surface and a confining substrate (pore  surface or other crystals)~\cite{Scherer1999,Flatt2002,Steiger2005a}.
While the thermodynamic equilibrium description provides fundamental
understanding of the forces at play in this problem~\cite{Steiger2005a}, 
the non-equilibrium kinetics of 
the relevant transport processes induce additional complexity
via the combined effects of attachment-detachment kinetics 
at crystal-solution interface, diffusion 
of solute crystal ions or molecules, and convection.

Following the seminal phenomenological approach by Weyl~\cite{Weyl1959},
some theoretical efforts have been devoted 
to the description of crystal growth and dissolution 
in grain-grain or grain-pore contacts~\cite{Coble1963,Elliott1973,Rutter1976,Raj1982,Kruzhanov1998,Fowler1999}. 
However, one systematic limitation of these studies is the lack
of description of microscopic interactions between 
the solid surface and the substrate, which are usually described via the 
disjoining pressure in thin films~\cite{Israelachvili1991}.
These interactions  combine with the spontaneous remodeling of the surface
via dissolution and growth to determine the contact morphology and dynamics.
In the past decades, Surface Force Apparatus (SFA) have allowed one
to probe interactions between surfaces at the nanoscale. 
These experiments have provided quantitative tests for 
standard theories such as the DLVO approach~\cite{Overbeek1947,Israelachvili1991}
combining an exponential electrostatic repulsion
with a power-law Van der Waals attraction.
However, if on the one hand double layer repulsion 
is considered as the prototypical interaction, 
on the other hand it was found to be accurate 
at all separations only for smooth crystalline 
surfaces in dilute electrolyte 
solutions\cite{Alcantar2003}. 

For other surfaces and solutions, 
significant deviation especially at short range (few nanometers),
 were measured \cite{Veeramasuneni1997,Alcantar2003,Drummond2002,Dishon2009,Diao2016}.
These non-DLVO contributions to the interaction have been 
found to depend on the specific nature of the surfaces, 
the solvent, the ions in the solvent, 
and the ions adsorbed on the surfaces. 
Using SFA, or atomic force microscope (AFM) on systems relevant to pressure solution
such as silica compounds (mica, silica colloids) and soluble salts, 
different authors have revealed the existence 
of additional repulsive interactions at short distances 
(a few nanometers) referred to as hydration forces~\cite{Alcantar2003,Delgado2005,Hamilton2010}.
These interactions,
the exact mathematical form of which is still matter of debate, are often
recognized to be exponentially decaying.
Beyond hydration forces, other specific interactions
include~\cite{Israelachvili1991}
oscillations at the molecular scale due to liquid ordering,
solute induced effects, depletion effects, etc. 
Owing to this wide variety of behaviors, we aim at 
developing an approach which is able to relate
the form of the interaction potential and
the dynamics during pressure solution, or growth. In this paper,
we focus on generic repulsive interaction potential
--such as exponential or power-law.
Our first goal is to question the role of the form of the
interaction potential on the dynamics of pressure solution.

A second goal is to identify the consequences 
of hydrodynamic convection in the thin liquid film. 
Indeed, convection has long been recognized 
to be important for solids growing with unconstrained interfaces, 
both in dendritic growth 
arising from solidification~\cite{Tonhardt1998}, and in growth from a solution~\cite{Wilcox1983}. 
However, its consequences have not been discussed in confined geometries.
In the absence of dissolution or growth, 
the hydrodynamics of squeezed films have been extensively
studied in the literature. This is known to
lead to an  evolution
of the thickness of the film exhibiting a non-trivial dependence 
on the solid geometry and  dimensionality~\cite{nasa_report1991}. 
In pressure solution, the geometry of the dissolving surface evolves in time, 
and emerges from a coupling between different forces 
and mass transport processes at play in the system. 
However, a complete description of growth with hydrodynamics 
(see e.g.~\onlinecite{Medvedev2005,Hammer2008,Tao2016}),
requires considerable numerical effort since it involves 
the concomitant solutions of the three-dimensional 
Navier-Stokes equation and of the evolution of the morphology of the crystal-liquid interface.

Here we propose to tackle this problem accounting 
consistently for thermodynamics, 
interaction effects (i.e. disjoining pressure), 
and non-equilibrium transport processes including diffusion 
and convection within a thin film approach which exploits 
the natural geometric slenderness of the contact region via the lubrication approximation~\cite{Oron1997}.
This method leads to a reduction of dimensionality, 
thereby facilitating numerical and analytical investigations.

The first part this paper in Sec.~\ref{sec:model} presents 
a three-dimensional continuum model which takes into account 
dissolution or growth, disjoining pressure effects, 
diffusion and hydrodynamics. 
The key assumption that the film is thin in the contact region 
is then formalized with the help of a multi-scale
expansion defining the lubrication limit~\cite{Oron1997}. 
This limit, widely employed in engineering (trust bearing)~\cite{nasa_report1991}, 
physics (nanoscale dewetting)~\cite{Karis2005,Matar2005} and biophysical models (membranes) ~\cite{Blount2015,LeGoff2014}, 
results in nonlinear and nonlocal thin film
evolution equations for the profile of the crystal surface. 
The end of \cref{sec:model} presents equations
for pressure solution in single contacts with
some simplifying assumptions such as equal densities
between the liquid and the solid, imposed symmetry
(left-right symmetric ridge or axisymmetric contact),
and dilute limit. 

Section~\ref{sec:methods} is devoted to the discussion
of relevant dimensionless numbers, and numerical methods.

In Sec.~\ref{sec:results}, we focus on the analysis 
of pressure solution for a single contact.
We investigate steady-states with
a time-independent surface profile and a fixed contact area.
We consider
 two different classes of repulsive interactions 
between the crystal surface and the substrate: divergent at contact and finite at contact. 

The dissolution rate is found to increase indefinitely with increasing load
in the case of diverging repulsions.  Viscosity effects
then become relevant for large enough loads. 
However, in the case of finite repulsions,
the dissolution rate is independent 
both on the viscosity and on the load
at large loads.

Moreover, as expected intuitively, the shape of the solid is flattened in the contact region for diverging 
repulsions. However, we find sharp and pointy contact shapes for finite repulsions.
In the limit of large loads, surface tension is found to be irrelevant
for diverging repulsions, while it is crucial in the case of finite repulsions
to regularize the pointy shapes at small scales.

We have also investigated the 
effect of dimensionality via the comparison of one-dimensional ridge contacts,
and two-dimensional axisymmetric contacts.
Dimensionality does not induce any qualitative
change in the behavior of pressure solution for diverging repulsions.
However, for finite repulsions and when surface tension
is neglected, the minimum distance between the dissolving solid and the 
substrate decreases exponentially with the load
in the ridge geometry, while it reaches zero for 
a finite force in the axisymmetric case.
Surface tension then comes into play at large enough loads,
and forbids real contact in the axisymmetric geometry.

Finally, the results are summarized and discussed in Sec.\ref{sec:discussion}.

\section{Model equations}
\label{sec:model}

\subsection{Dissolution and Growth in a liquid}

\begin{figure}[t]
\includegraphics[width=\linewidth]{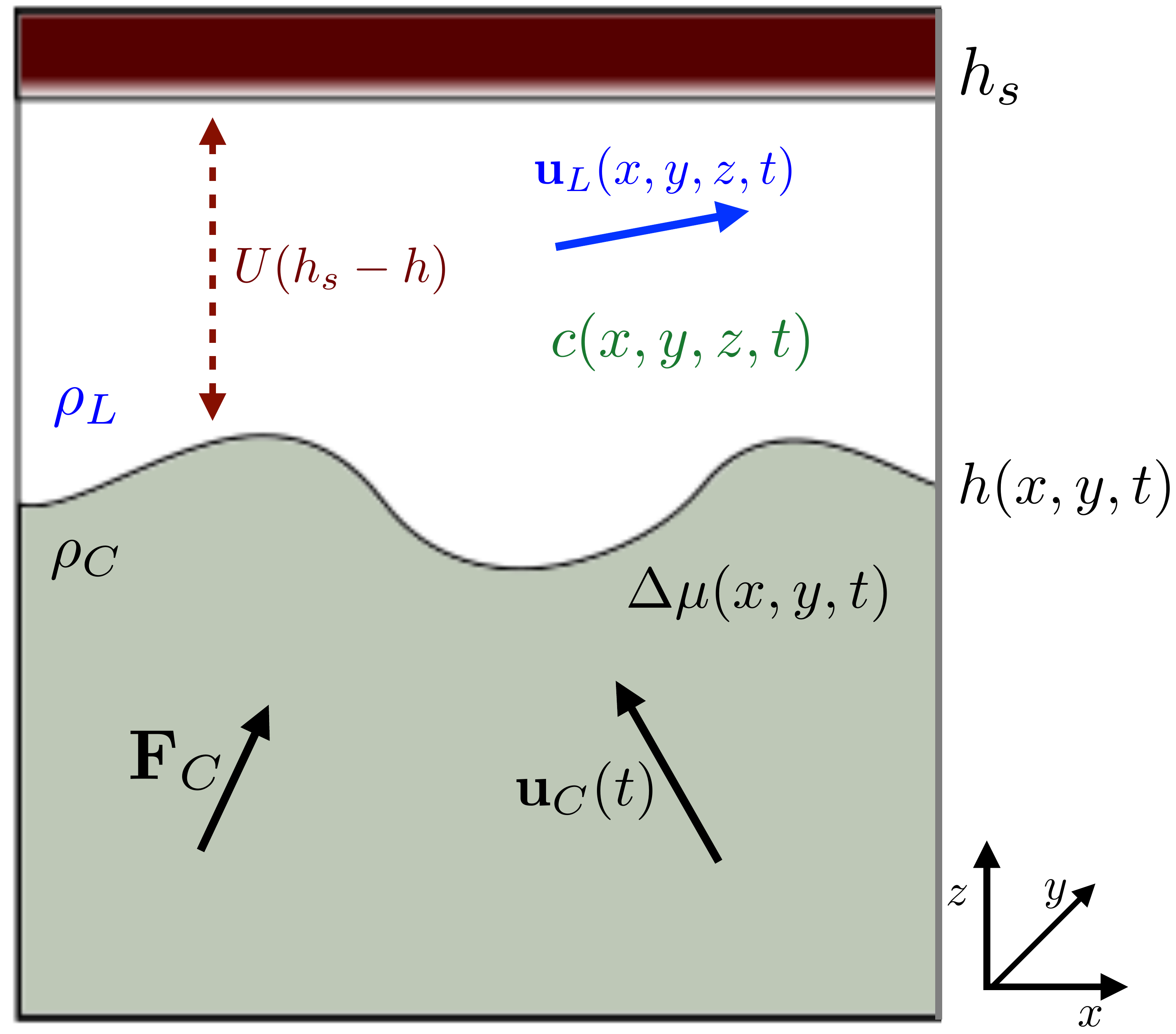}
\caption{
Sketch of the contact region with some variables and fields
of the model. See text for notations.}
\label{fig:model_sketch_zoom}
\end{figure}

The system under study is represented in \cref{fig:model_sketch,fig:model_sketch_zoom}.
For the sake of clarity, 
we designate the growing
or dissolving solid by the name crystal.
However, our model equally applies to amorphous phases,
or to any other solid phases that can grow
and dissolve. We consider a crystal in a liquid medium, 
growing or dissolving in the vicinity of a substrate,
and subjected to an external force or load $\mathbf{F}_{C}(t)$.
The crystal is assumed to be rigid, namely we neglect the contribution 
of elastic deformations on the interface shape and chemical potential. 
For the sake of simplicity, 
we also discard crystal rotations and consider only translations.
The substrate  at $z=h_s(x,y)$ is immobile i.e. $\partial_t h_s = 0$, 
 and is impermeable. 
The liquid crystal interface (LC) at $z=h(x,y,t)$ evolves with time.

We assume an incompressible fluid with constant density $\rho_L$ 
\begin{equation}
\label{eq:Newtonian}
\nabla \cdot \mathbf{u}_L  =0 \, .
\end{equation} 
Neglecting inertial effects (which are known to be negligible
in the lubrication limit considered below~\cite{Guyon2015,Oron1997}), 
the liquid obeys the Stokes equation:
\begin{equation}
\label{eq:Navier_Stokes}
\eta \nabla^2 \mathbf{u}_L  = -\nabla p \, ,
\end{equation} 
where $\eta$ is the viscosity,
and $p(x,y,z,t)$ is the pressure.
Global mass conservation at the LC interface (neglecting 
possible mass excess at the interface) reads~\cite{Wilcox1983}
\begin{equation}
\label{eq:mass_global}
\rho_L (\mathbf{u}_L\cdot \hat{\mathbf{n}} - v_n) = \rho_C (\mathbf{u}_C\cdot \hat{\mathbf{n}} - v_n)\, ,
\end{equation}
where $\rho_C$ is the constant crystal density, 
$\mathbf u_C$ is the translational velocity of the rigid crystal,
$\hat{\mathbf n}$ is the normal to the LC interface
and $v_n$ is the normal velocity of the interface.
Note that whenever a three-dimensional field
such as  $\mathbf u_L$  appears in an equation 
evaluated at an interface, we consider implicitly the value
of this field at this interface.
Finally, we assume no slip and no penetrability  at the liquid-substrate (LS) interface
\begin{equation}
\label{eq:boundary_hs}
\mathbf{u}_L = \mathbf 0\, ,
\end{equation}
and a no slip condition at the LC interface
\begin{equation}
\label{eq:boundary_h}
\mathbf{u}_{L\parallel} = \mathbf{u}_{C\parallel}\, ,
\end{equation}
where the index $\parallel$ indicates the projection 
of a vector on
the plane tangent to the LC interface. 

Local mass conservation of the solute (crystal ions or molecules in the fluid) 
reads in the liquid bulk
\begin{equation}
\label{eq:Fick}
\partial_t c + \mathbf{u}_L \cdot \nabla c = -\nabla\cdot \mathbf{j}\, ,
\end{equation}
where $\mathbf{j}$ is the diffusion flux. 
We assume that diffusion is governed by  Fick's law 
\begin{eqnarray}
\label{eq:ficks_law}
\mathbf{j} = -D(c)\nabla c.
\end{eqnarray}
At the LC interface solute mass conservation imposes
\begin{equation}
\label{eq:mass_crystal}
\Omega^{-1}(v_n-\hat{\mathbf{n}}\cdot \mathbf{u}_C) 
= c (v_n - \hat{\mathbf{n}}\cdot \mathbf{u}_L) - \hat{\mathbf{n}}\cdot \mathbf{j}\, ,
\end{equation}
where  $\Omega$ is the molecular volume in the crystal.

Assuming that the substrate is impermeable at the LS interface, we have
\begin{equation}
\label{eq:mass_substrate}
\mathbf{j}\cdot \hat{\mathbf{n}}_s = 0\, ,
\end{equation}
with $\hat{\mathbf{n}}_s$ the LS interface normal.

The crystallization/dissolution rate
$v_n - \mathbf{n}\cdot \mathbf{u}_C$
is assumed to depend linearly on the departure from equilibrium
\begin{equation}
\label{eq:linear_kinetics}
v_n - \mathbf{n}\cdot \mathbf{u}_C = \Omega\nu (c-c_{eq})\, ,
\end{equation}
where $\nu$ is a kinetic coefficient and $c_{eq}(x,y,t)$ 
the local equilibrium concentration. In the ideal limit,
where the activity coefficient is equal to $1$, we have
\begin{equation}
\label{eq:concentration}
c_{eq} =c_0 e^{{\Delta \mu}/{k_BT}}\, ,
\end{equation}
where $\Delta \mu$ is the local chemical potential 
of the crystal at the interface 
and $c_0$ is the equilibrium concentration for an interface 
in an infinitely large crystal far from the substrate (solubility).
The chemical potential at the LC interface reads
\begin{equation}
\label{eq:mu}
\frac{\Delta \mu(x,y,t)}{\Omega}  =\tilde{\gamma}: \kappa
+ W'(x,y,h) + (\frac{\rho_C}{\rho_L}-1)\sigma_{nn}\, ,
\end{equation} 
where $\tilde{\gamma} (x,y)$ is the stiffness tensor~\cite{Saito1996}, 
$\kappa$ is the curvature tensor, 
$W'=\partial_hW(x,y,h)$ is
the disjoining pressure~\cite{Quere2003}.
The potential $W(x,y,h)$ is taken to depend on $x$ and $y$
to account for the possible spatial heterogeneities 
of the substrate height  $h_s$, and 
 of the substrate material properties. 
The liquid stress tensor is defined as $\sigma = \sigma' - \delta_{ij} p$ 
with $\sigma'_{ij} = \eta(\partial_j u_{Li} +\partial_i u_{Lj})$,
and the index $n$ indicates the normal direction.
The last term of \cref{eq:mu} accounts for the energy cost 
associated to the volume change during the phase transformation.

Finally, since the crystal is a rigid body, and since we neglect inertia,
we write a global force balance  on the crystal as
\begin{equation}
\label{eq:force_balance_general}
\mathbf{F}_{C} = \oiint_{LC} \!\mathrm{d}S\,  
[ -\hat{\bf n}\cdot\sigma +\hat{\bf n}(\tilde{\gamma}: \kappa+ W')]\, ,
\end{equation}
where the surface integral is performed along all the LC  interface
(since we discard crystal rotations, we do not consider the equilibrium of torques).

The system of equations reported above
describes the dissolution or growth dynamics of a rigid crystal interacting with a 
frozen and impermeable substrate. In the following, we specialize the discussion
to the contact region.


\subsection{Contact region}

In this section, we re-write mass conservation and force balance
in a form which makes use of the geometry of the contact region.
We assume that the LS and LC interfaces exhibit no overhang. 

For any field $g(x,y,z)$ defined everywhere in the liquid, we consider the  following integrated quantity along $z$
\begin{equation}
\label{eq:planar_current}
 \langle g\rangle(x,y) = \int_{h(x,y)}^{h_s(x,y)}\mkern-36mu \mathrm{d}z\, g(x,y,z).
\end{equation}
Using the incompressibility condition \cref{eq:Newtonian}, 
the immobility of the substrate \cref{eq:boundary_hs}, 
and global mass conservation at the LC interface \cref{eq:mass_global}, we 
obtain a two-dimensional equation for liquid mass conservation
\begin{align}
\label{eq:mass_conservation}
\frac{\rho_C}{\rho_L}v_{Cz}
= -\nabla_{xy}\cdot \langle\mathbf{u}_{Lxy}\rangle -\partial_t (h_s-h)\, .
\end{align}
Here and in the following,
vectors with the index $_{xy}$ indicate the two-dimensional vector 
in the $x,y$ plane without the $z$ component.
In addition, we have used
the geometric relations
\begin{align}
\hat{\mathbf{n}} &= \frac{(-\nabla_{xy}h,1)}{[1+(\nabla_{xy}h)^2]^{{1}/{2}}}\, ,\\
v_n &=\frac{\partial_t h}{[1+(\nabla_{xy}h)^2]^{{1}/{2}}}\, ,
\end{align}
and we have defined the crystallization-dissolution rate along $z$
\begin{eqnarray}
v_{Cz}
=\partial_th-u_{Cz}+\mathbf u_{Cxy}\cdot \nabla_{xy} h\, .
\label{e:definition_vCz}
\end{eqnarray}

Similarly, using \cref{eq:mass_substrate,eq:mass_crystal},
mass conservation for
the solute concentration $c$ can be re-written in a two-dimensional form
\begin{equation}
\label{eq:crystal_conservation}
\frac{v_{Cz}}{\Omega}
+ \partial_t  \langle c\rangle
+ \nabla_{xy}\cdot \langle\mathbf{u}_{Lxy}c\rangle
= -\nabla_{xy}\cdot \langle\mathbf{j}_{xy}\rangle \, .
\end{equation}

In order to write the force balance at the contact,
we make use of two additional physical assumptions.
First, we assume that the pressure outside the contact
is approximately constant and equal to $p^{ext}$.
Second, we assume that the interaction term
vanishes, i.e. $W'\approx 0$, away from the contact.

Finally, assuming that $\sigma'$ vanishes outside the contact region force balance  \cref{eq:force_balance_general} is re-written as 
\begin{equation}
\label{eq:force_balance}
\mathbf{F}_C = \iint_{\text{contact}} \mkern-36mu\mathrm{d}S\, 
[\hat{\bf {n}}(p-p^{ext}+W'(x,y,h)) -\hat{\bf {n}}\cdot\sigma' ]\, .
\end{equation}
This equation makes use of fact that 
the total force exerted by surface tension
or by a constant external pressure
on a crystal of arbitrary shape vanishes
\begin{eqnarray}
\oiint_{LC} \!\mathrm{d}S\, \hat{\bf n}( \tilde{\gamma} : \kappa) &=&0.
\nonumber\\
\oiint_{LC} \!\mathrm{d}S\, \hat{\bf n}\, p^{ext}&=&0.
\end{eqnarray}
These two identities are proved in \cref{appendix:force_balance}.

\subsection{Lubrication limit in the contact region}
\label{section:lubrication_limit}

Here, we show that  lubrication limit
based on the small slope approximation,
allows one to express the quantities integrated along $z$ in \cref{eq:planar_current,eq:crystal_conservation,eq:force_balance},
thereby leading to closed-form equations for three quantities.
The two first quantities are time and space-dependent
fields: the pressure $p$, the thickness of the liquid film
\begin{align}
\label{e:zeta}
\zeta(x,y,t)=h_s(x,y)-h(x,y,t).
\end{align} 
The third quantity is the time-dependent crystal velocity $\mathbf u_{C}$.

The lubrication limit~\cite{Guyon2015} makes use 
of a disparity of length scales: 
the lateral extent of the film is assumed to be large $x \sim\mathcal{O}(\epsilon^{-1})$ 
as compared to the film thickness $\zeta=(h_s-h)\sim \mathcal{O}(1)$ with $\epsilon\ll 1$.
The mathematical procedure to derive these equations is
well known~\cite{Guyon2015,Oron1997}, and we therefore only provide the main steps
of the derivation.
Formally, we identify a small parameter $\epsilon = h_0/l$, 
where $l$ is the typical extent of the contact region 
and $h_0$ is the typical gap between the crystal and the substrate.
Spatial coordinates then scale 
with this small parameter as
$x\sim y \sim \ell\sim  h_0/\epsilon$, and $z\sim h_0$.
Furthermore, assuming that the typical fluid velocity
parallel to the substrate is $\mathbf{u}_{Lxy}\sim u_0$, we also consistently choose
$u_{Lz}\sim \epsilon u_0$, 
pressure $p \sim \eta u_0 /(\epsilon h_0)$, and time $t\sim h_0/\epsilon u_0$.

Substituting these scalings of physical variables
 in the model equations 
we obtain the lubrication expansion~\cite{Guyon2015,Oron1997}.
To leading order, \cref{eq:Navier_Stokes} reduces to
\begin{align}
\label{e:lubric_dpdz}
\partial_z p  =0\, ,\\
-\nabla_{xy} p +\eta \partial_z ^2 \mathbf{u}_{Lxy}  =0\, .
\end{align}
The first equation indicates that the 
pressure does not depend on $z$, but only on $x,y$, and $t$.
Solving the second equation using the boundary conditions \cref{eq:boundary_h,eq:boundary_hs} 
results in a Poiseuille (parabolic) flow for $\mathbf{u}_{xy}$,
\begin{equation}
\label{eq:liquid_vel_lubri}
\mathbf{u}_{Lxy} = -\frac{(h_s-z)(z-h)}{2\eta}\nabla_{xy} p + \frac{h_s-z}{\zeta} u_{Cxy}\, .
\end{equation}
Integrating over the film thickness, we obtain
\begin{equation}
\label{eq:flux_lubri}
\langle\mathbf{u}_{Lxy}\rangle= -\frac{\zeta^3}{12\eta} \nabla_{xy} p + \frac{\zeta}{2}\mathbf{u}_{Cxy}\, .
\end{equation}
Combining \cref{eq:mass_conservation,eq:flux_lubri}, we obtain
\begin{flalign}
\frac{\rho_C}{\rho_L} v_{Cz}
= \nabla_{xy}\cdot\left[ \frac{\zeta^3}{12\eta}\nabla_{xy}p 
- \mathbf u_{Cxy}\frac{\zeta}{2}
\right]
-\partial_t\zeta.
\label{e:2D_mass_cons_hydro_p}
\end{flalign}

A similar procedure is applied to the concentration field.
Assuming $c\sim\mathcal{O}(1)$ we obtain to leading order from \cref{eq:Fick}
\begin{equation}
\partial_z [D(c)\partial_z c] = 0\, .
\end{equation}
Integrating this relation, and using local conservation of mass at the boundaries 
\cref{eq:mass_crystal,eq:mass_substrate}, we obtain $\partial_z c=0$,
showing that the concentration does not depend on $z$. 
Furthermore, assuming
finite attachment-detachment kinetics $\nu\sim\mathcal{O}(1)$
in \cref{eq:linear_kinetics}  we obtain:
\begin{align}
\label{eq:c_lubri}
c = c_{eq}(x,y,t)\, .
\end{align}
Hence for finite attachment-detachment kinetics, 
the concentration to leading order in the lubrication limit is equal
to the local equilibrium concentration. 
This is the consequence of the smallness the film thickness
which enforces slow diffusion along the film, leaving ample time
for local equilibration of the concentration via attachment and detachment
of the LC interface.
We may now write \cref{eq:crystal_conservation} 
using \cref{eq:ficks_law} in the lubrication limit as
\begin{flalign}
\frac{v_{Cz}}{\Omega}
+\partial_t[\zeta c_{eq}]
-\nabla_{xy}\cdot\left[\frac{\zeta^3}{12\eta}c_{eq}\nabla_{xy}p \right]
+\frac{\mathbf u_{Cxy}}{2}\cdot\nabla_{xy}[c_{eq}\zeta]
\nonumber \\
= \nabla_{xy} \cdot [\zeta{D(c_{eq})} \nabla_{xy} c_{eq}].
\label{e:2D_mass_cons_c_lubric_fastkin_p}
\end{flalign}

This relation involves $c_{eq}$, which depends 
on the chemical potential via \cref{eq:concentration}.
Let us compare the different contributions of the 
chemical potential.
The lubrication expansion  imposes $p\sim\mathcal{O}(\epsilon^{-1})$.
For disjoining forces to be able to balance
viscous forces, we choose $W'(x,y,h)\sim  \mathcal{O}(\epsilon^{-1})$.
As a consequence, the pressure term and the interaction term 
in \cref{eq:mu} are
of the same order of magnitude. 
In addition, since the curvature 
$\kappa\sim \partial_{xx}h\sim\partial_{yy}h\sim \epsilon^{2}$ is small,
only large stiffnesses $\tilde\gamma\sim   \mathcal{O}(\epsilon^{-3})$
can make the capillary term $\tilde\gamma:\kappa$ relevant. However
even if surface stiffness is not so large, 
the capillary term can be relevant in two cases:
(i) if the curvature locally blows up,
and (ii) far from the substrate where the potential term $W'$ can be neglected.
We will see in the following  that these conditions 
can be reached during pressure solution.
Therefore, in order to include all relevant cases for
the discussion below, we keep the capillary term
leading to
\begin{align}
\label{eq:mu_lubrication}
\frac{\Delta \mu(x,y,t)}{\Omega}  &=-\tilde{\gamma}_1\partial_{x_1x_1}h-\tilde{\gamma}_2\partial_{x_2x_2}h
\nonumber \\
&+ W'(x,y,h) + (\frac{\rho_C}{\rho_L}-1)p\, ,
\end{align} 
where $x_1$ and $x_2$ are the directions of principal curvature
of the LC interface, and $\tilde{\gamma}_1,\tilde{\gamma}_2$
are the related surface stiffnesses~\cite{Saito1996}.

Finally, since $W'$ is of the same order
as $p$ in the lubrication limit,
force balance \cref{eq:force_balance} reads
\begin{align}
\label{eq:force_balance_lubri}
F_{Cz} &= \iint_{\text{contact}}\mkern-30mu \mathrm{d}A\, (p-p^{ext} +W'(x,y,h))\, ,\\
\label{eq:force_balance_lubri_xy}
\mathbf F_{Cxy} &=\iint_{\text{contact}} \mkern-30mu \mathrm{d}A\, 
\left(\frac{\eta \mathbf u_{Cxy}}{\zeta} 
- (p-p^{ext})\nabla_{xy}(h_s-\frac{\zeta}{2})\right)\, ,
\end{align}
where $\mathrm{d}A = \mathrm{d}x\,\mathrm{d}y$. 
To derive the last relation,
we have assumed that, at the boundary of the contact zone, $p=p^{ext}$
is constant and $\zeta$ is large enough for $W'$ to be negligible.

As a summary, we have derived a thin film model for the contact region  
during dissolution and growth, which consists of two equations 
\cref{e:2D_mass_cons_hydro_p,e:2D_mass_cons_c_lubric_fastkin_p}
for the coupled two-dimensional space and time dependent
fields $p$ and $\zeta$, and an additional vectorial
integral constraint \cref{eq:force_balance_lubri,eq:force_balance_lubri_xy} 
which determines the time-dependent crystal velocity $\mathbf u_C$.
This system is not only nonlinear, but also nonlocal due to the 
force balance equation. 
In the following, we explore some consequences of the model 
in the specific case of pressure solution of
a single contact.

\subsection{Ridge and axisymmetric contact}

We now consider the pressure solution of a single contact
with some simplifying assumptions:
\begin{itemize}
\item
(i) equal densities 
between the liquid and the crystal $\rho_C=\rho_L$;
\item
(ii) no lateral motion  $\mathbf{u}_{Cxy} =0$ and no lateral force  $F_{Cxy}=0$;
\item
(iii) diffusion constant independent of concentration $D(c) = D$;
\item
(iv) isotropic surface tension  
$\tilde\gamma_1=\tilde\gamma_2=\gamma$;
\item
(v) flat substrate $h_s$ independent of $x$ and $y$. We use the interaction potential $U$, defined by
$U(\zeta)=W(x,y,h)$. It follows that $W'(h)=\partial_hW(h)=-\partial_\zeta U(\zeta)=-U'(\zeta)$.
\item
(vi) small concentrations $\Omega c_{eq}\ll 1$;
\item
(vii) linearised Gibbs-Thomson relation $\Delta \mu/k_BT\ll 1$.
\end{itemize} 

In addition, we consider two simple geometries.
The first one is a one-dimensional ridge, which is invariant
along $y$, and left-right symmetric with $h(x)=h(-x)$.
The second geometry is an axisymmetric contact, the shape of which
depends only on the distance $r$ from the origin in the $x,y$ plane.
In the following we will often refer to the symmetric  ridge as 1D, 
and the axisymmetric contact as 2D.

\subsubsection{Symmetric ridge}

Consider first the ridge case obeying the 
$x\rightarrow -x$ symmetry, with a 
system length $2L$.  Assuming $\rho_C=\rho_L$,
the integration of \cref{e:2D_mass_cons_hydro_p} leads to
\begin{equation}
\label{eq:pressure}
p = p^{ext} + u_{Cz}\int_x^{L} \mathrm{d}x\, \frac{12\eta x}{\zeta^3}\, .
\end{equation}
Plugging this expression into \cref{eq:force_balance_lubri} provides us 
with a non local relation between the crystal velocity and the surface height:
\begin{equation}
\label{eq:vel_1d}
2u_{Cz}  \int_0^L \mathrm{d}x\int_x^L \mathrm{d}x' \frac{12\eta x'}{\zeta^3} 
= F_{Cz}^{1D} + 2\int_0^L \mathrm{d}x\, U'(\zeta)\, .
\end{equation}

This equation relates the sum of the load and 
interaction forces between the crystal and the substrate
on the right hand side, to the forces caused by viscous dissipation
in the film on the left hand side. In the viscous term, the crystal velocity $u_{Cz}$
is multiplied by the hydrodynamic mobility of the crystal which depends on the 
interface profile $\zeta$.  The expression of this mobility
is well known in the lubrication limit\cite{nasa_report1991}.

In the limit of small concentrations $\Omega c_{eq} \ll 1$
and equal densities $\rho_L=\rho_C$, \cref{e:2D_mass_cons_c_lubric_fastkin_p}
takes a simple form
\begin{equation}
\label{eq:height_implicit}
\partial_t \zeta = -D\Omega\partial_x[\zeta\partial_x c_{eq}] - u_{Cz} \, .
\end{equation}
Assuming that $\Delta \mu/k_BT\ll 1$ in  \cref{eq:concentration} and
using \cref{eq:mu_lubrication},
we obtain
\begin{equation}
\label{eq:height_1d}
\partial_t \zeta = -D_{e}\partial_x\Bigl[\zeta  \partial_x (\gamma \partial_{xx} \zeta - U'(\zeta))\Bigr] - u_{Cz}\, ,
\end{equation}
where by definition
\begin{equation}
D_{e} = \frac{D\Omega ^2 c_0}{k_BT}.
\end{equation}

\subsubsection{Axisymmetric contact}
Let us now consider an axisymmetric contact. Using cylindrical coordinates in a contact zone of radius $R$, we obtain in a similar way the following equations
\begin{align}
\label{eq:vel_2d}
&2u_{Cz} \, \pi\int_0^R \mathrm{d}r\, r\int_r^{R} \mathrm{d}r'\, \frac{6\eta r'}{\zeta(r')^3} 
= F_{Cz}^{2D} + 2\pi\int_0^R \mathrm{d}r \, r U'(\zeta)\, ,\\
\label{eq:height_2d}
&\partial_t \zeta = 
-D_{e}\frac{1}{r}\partial_r\Bigl[r\zeta \partial_r(\gamma \partial_{rr} \zeta 
+\frac{\gamma}{r}\partial_r \zeta  - U'(\zeta)) \Bigr] - u_{Cz}\, ,
\end{align}
where the quantity proportional to $\gamma$ is the mean curvature 
in axial symmetry \cite{Weatherburn1955,Goldman2005}.

\subsubsection{Interaction potentials}

We chose to study two generic types of repulsive interaction 
potentials.
The first one diverges when the film thickness $\zeta$ vanishes
\begin{equation}
\label{eq:singular_pot}
U(\zeta) = \frac{A}{\zeta^n}\, ,
\end{equation}
where $A$ is a constant.
In practice numerical results have been
obtained with $n=3$. However we will keep an arbitrary 
exponent $n$ in the discussions. 

The second type of potential exhibits a finite repulsion 
when $\zeta\rightarrow 0$ 
\begin{equation}
\label{eq:interaction_exp}
U(\zeta) = Ae^{-\frac{\zeta}{\lambda}}\, ,
\end{equation}
where $\lambda$ is a decay length
representing for instance the Debye length in the case of electrostatic
interactions~\cite{Israelachvili1991}.

The essential
difference between these potentials is that \cref{eq:singular_pot}
leads to an infinite repulsion force when $\zeta\rightarrow 0$,
whereas this force is finite for \cref{eq:interaction_exp}.

\section{Methods}
\label{sec:methods}

\subsection{Normalization}
\label{sec:normalization}

In order to perform simulations and to analyze the 
results of the model, we write the model equations
in a dimensionless form and identify the relevant
dimensionless parameters. 
All variables appearing in normalized units are labeled with a top bar.

We start by defining the dimensionless repulsion strength $\bar{A}$.
For the exponential potential we set $\bar{A}=A/\gamma$, 
while for power-law repulsions with the case $n=3$, we use $\bar{A}=A/(\gamma\lambda^3)$.
The normalized film thickness is $\bar \zeta=\zeta /\lambda$,
and the normalized coordinates are $\bar x=x \bar A^{1/2}/\lambda$,
$\bar y=y \bar A^{1/2}/\lambda$. 
The normalized time is defined as 
$\bar t= t D_e \gamma {\bar A}^2/\lambda^3$.
The normalized equations are showed in \cref{appendix:scaling}. 
Notice that the scale $\lambda$ is imposed
by the expression of $U$ in the case of
an exponential repulsion, while
it is an arbitrary lengthscale corresponding
to the actual film width in the case
of power-law repulsions.

The normalized repulsion strength  $\bar{A}$
comes into play in spatio-temporal scales 
but not as a parameter of the normalized
equations. As a consequence, it cannot change
the model behavior qualitatively.
The only parameters explicitly appearing in the 
normalized equations are the normalized 
viscosity $\bar{\eta}$, and external load $\bar{F}_{Cz}$.
The normalized viscosity reads
\begin{align*}
\bar{\eta} =\frac{D_{e}}{ \lambda^2}\eta=\frac{D\Omega^2c_0}{ \lambda^2k_BT}\eta\, .
\end{align*}
Since the loads have different dimensionality in 1D (force per unit length) and 2D (force),
their normalization is different
\begin{align*}
\bar{F}^{1D}_{Cz}=\frac{{F}^{1D}_{Cz}}{\gamma \bar A^{1/2}}\, ,
\\
\bar{F}^{2D}_{Cz}=\frac{{F}^{2D}_{Cz}}{\gamma \lambda}\, .
\end{align*}

Below, all simulations are performed with normalized
variables and coordinates. However, the analysis 
of the equations is performed in physical coordinates
to make the physical interpretation more transparent.

\subsection{Numerical methods}
\label{sec:num_methods}

We solved \cref{eq:height_1d,eq:vel_1d} or \cref{eq:height_2d,eq:vel_2d} 
using an explicit Euler method, where derivatives
are calculated with the help of a finite difference scheme. 
We imposed a fixed interface height at the boundary of the contact region, 
$\zeta= \zeta_{bc}$ where $x=\pm L$ or $r=R$.
The gap $\zeta_{bc}$ between the crystal and the substrate
at the boundary is chosen to be large as compared 
to the range of the interaction potential, but small as compared
to the contact region width $L$, or $R$. 
We also impose a constant supersaturation at the boundary 
$\Delta C = c_{eq}/c_0-1 \approx \Delta \mu/(k_BT)$ 
to mimic a macroscopic concentration bath outside
the contact.

The boundary conditions introduce three additional 
dimensionless parameters.
The normalized system size
\begin{eqnarray}
\bar L= \frac{L\bar A^{1/2}}{\lambda},
\hspace{0.5 cm}{\text or}\hspace{0.5 cm}
\bar R =\frac{R\bar A^{1/2}}{\lambda},
\end{eqnarray}
the normalized film thickness at the boundary
\begin{eqnarray}
\bar \zeta_{bc}= \frac{\zeta_{bc}}{\lambda},
\end{eqnarray}
and the normalized supersaturation
\begin{eqnarray}
\overline{\Delta C}= \frac{k_BT\lambda}{\bar{A}\gamma\Omega}\Delta C .
\end{eqnarray}

Simulations are performed with $\bar{L},\bar{R}=100$, 
substrate position $\bar{h}_s=2$, film thickness at the boundary $\bar{\zeta}_{bc} =12$, 
and boundary supersaturation $\Delta C =0$.
The discretization bin size is $\Delta \bar{x} = 0.2$ for most simulations.
However in some cases, to be able to resolve the contact shape 
at very high external forces (see \cref{sec:exp_rep,fig:exp_shape}), 
it was necessary to increase the spatial resolution up to 16 times.

The simulations were always started with a flat profile  
(see top panel of \cref{fig:evolution_colormap}).
When applying a concentration higher than the equilibrium one  at the boundary, 
we observe crystal growth: the crystal translates downward
by addition of growth units at the surface, and $u_{Cz}<0$. 
When applying an external load, 
$F_{Cz}$ with sign in the positive direction hence pushing
 the crystal towards the substrate, we observe dissolution, 
i.e. pressure solution and $u_{Cz}>0$.
The latter case is the main focus of this paper.

\section{Results: Single contact pressure solution}
\label{sec:results}

\begin{figure}
\includegraphics[width=0.9\linewidth]{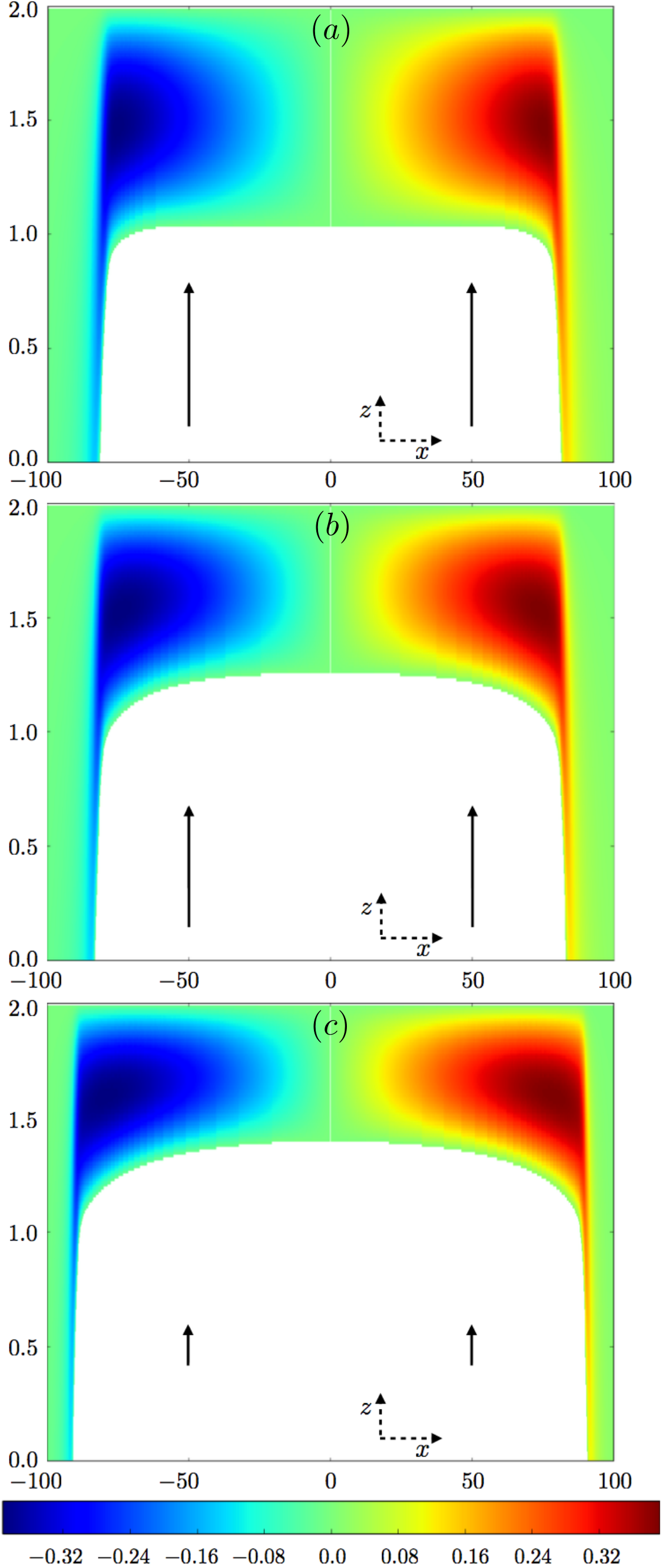}
\caption{Pressure solution dynamics.
Snapshots of the numerical solution of \cref{eq:height_1d} 
representing a dissolving contact ridge. Size of the simulation box $\bar{L} =100$ (physical size $L\approx 1\mathrm{\mu m}$) under an external pressure $p = 26$MPa. The viscosity is $\bar{\eta} = 0.5$.
The crystal is in white and the black arrows 
are proportional to the crystal velocity $u_{Cz}$. 
The time increases from the top panel to the bottom one.  
(a): initial condition.
As an example using physical constants related to 
calcite (see Sec.\cref{sec:order_magnitude}), physical time frames are: 
(b) 10s, (c) 6.7min. 
The colormap (in arbitrary units) shows the amplitude of 
the $x$ component of liquid velocity field $\mathbf u_{Lxy}$,
as obtained from \cref{eq:liquid_vel_lubri,eq:pressure}. 
The vertical scale is in nanometers. The substrate is located at $h_s = 2$nm.}

\label{fig:evolution_colormap}
\end{figure}

As an illustrative example
we show in \cref{fig:evolution_colormap}
the numerical solution for the profile of a ridge obeying \cref{eq:height_1d,eq:vel_1d} 
when an external load pushes the crystal upwards against the substrate, 
and when the interaction is in the form of a singular repulsion \cref{eq:singular_pot}. 
A similar shape is observed when solving \cref{eq:height_2d,eq:vel_2d} 
for an axisymmetric contact looking at the section along the radius. 
The simulation shows that the interface profile reaches a steady state characterized by a constant crystal 
velocity (dissolution rate) and fixed interface position.

As discussed earlier in \cref{section:lubrication_limit}, in the contact region and in the absence 
of blow-up of the curvature, we expect the surface tension contribution to be 
small. Neglecting this contribution, steady-state
solutions with a constant profile i.e. $\partial_t \zeta =0$,
obey respectively in 1D or 2D
\begin{subequations}
\label{eq:steady_state}
\begin{align}
0 = u_{cz} - D_{e}\partial_x [\zeta\partial_x U'(\zeta)]\, ,
\\
0 = u_{cz} - \frac{ D_e}{r}\partial_r [r\zeta\partial_r U'(\zeta)]\, .
\end{align}
\end{subequations}
This equation is integrated as
\begin{subequations}
\label{eq:general_largeForce}
\begin{align}
\frac{x^2}{2D_{e}}u_{Cz} = \tilde{U}(\zeta(r)) - \tilde{U}(\zeta_0) \, ,\label{eq:general_largeForce_1d}
\\
\frac{r^2}{4D_{e}}u_{Cz} = \tilde{U}(\zeta(r)) - \tilde{U}(\zeta_0) \, , \label{eq:general_largeForce_2d}
\end{align}
\end{subequations}
where $\zeta_0=\zeta(0)$, and $\tilde{U}(\zeta)$ is defined via the relation
\begin{equation}
\label{eq:Utilde_definition}
\tilde{U}'(\zeta) = \zeta U''(\zeta)\, ,
\end{equation}
which, up to an additive constant leads to $\tilde{U}(\zeta) = \zeta U'(\zeta)-U(\zeta)$.
Since we expect physically that the interaction potential tends to a constant as
$\zeta\rightarrow\infty$, i.e. that $U(\infty)$ is a  constant,
then $\tilde{U}(\infty)$ should also be a  constant.
Therefore, $\tilde{U}$ cannot increase indefinitely when $\zeta\rightarrow\infty$
on the r.h.s. of \cref{eq:general_largeForce_1d,eq:general_largeForce_2d}. As a consequence,
there are finite  $x_m$ or $r_m$ where $\zeta\rightarrow\infty$
and they obey
\begin{subequations}
\label{eq:general_largeForce_contact}
\begin{align}
\frac{x_m^2}{2D_{e}}u_{Cz} = \tilde{U}(\infty) - \tilde{U}(\zeta_0) \, ,\label{eq:general_largeForce_contact_1d}
\\
\frac{r_m^2}{4D_{e}}u_{Cz} = \tilde{U}(\infty) - \tilde{U}(\zeta_0) \,\label{eq:general_largeForce_contact_2d} .
\end{align}
\end{subequations}
Since $\zeta$ diverges at some finite distance $x_m$ or $r_m$
from the center of the contact, the size of the contact 
in steady-state pressure solution is always finite.

In the limit of large forces, we expect $\zeta_0$ to become small.
The situation then turns out to be very different
depending on how $\tilde{U}(\zeta_0)$ behaves when $\zeta_0$ is small.
The following sections discuss separately the cases
of finite and diverging interaction potentials $U(\zeta)$, corresponding
to finite or diverging $\tilde{U}(\zeta)$ as $\zeta\rightarrow 0$.

\subsection{Singular repulsion: power law case}
\label{sec:power_law}
\begin{figure}
\includegraphics[width=\linewidth]{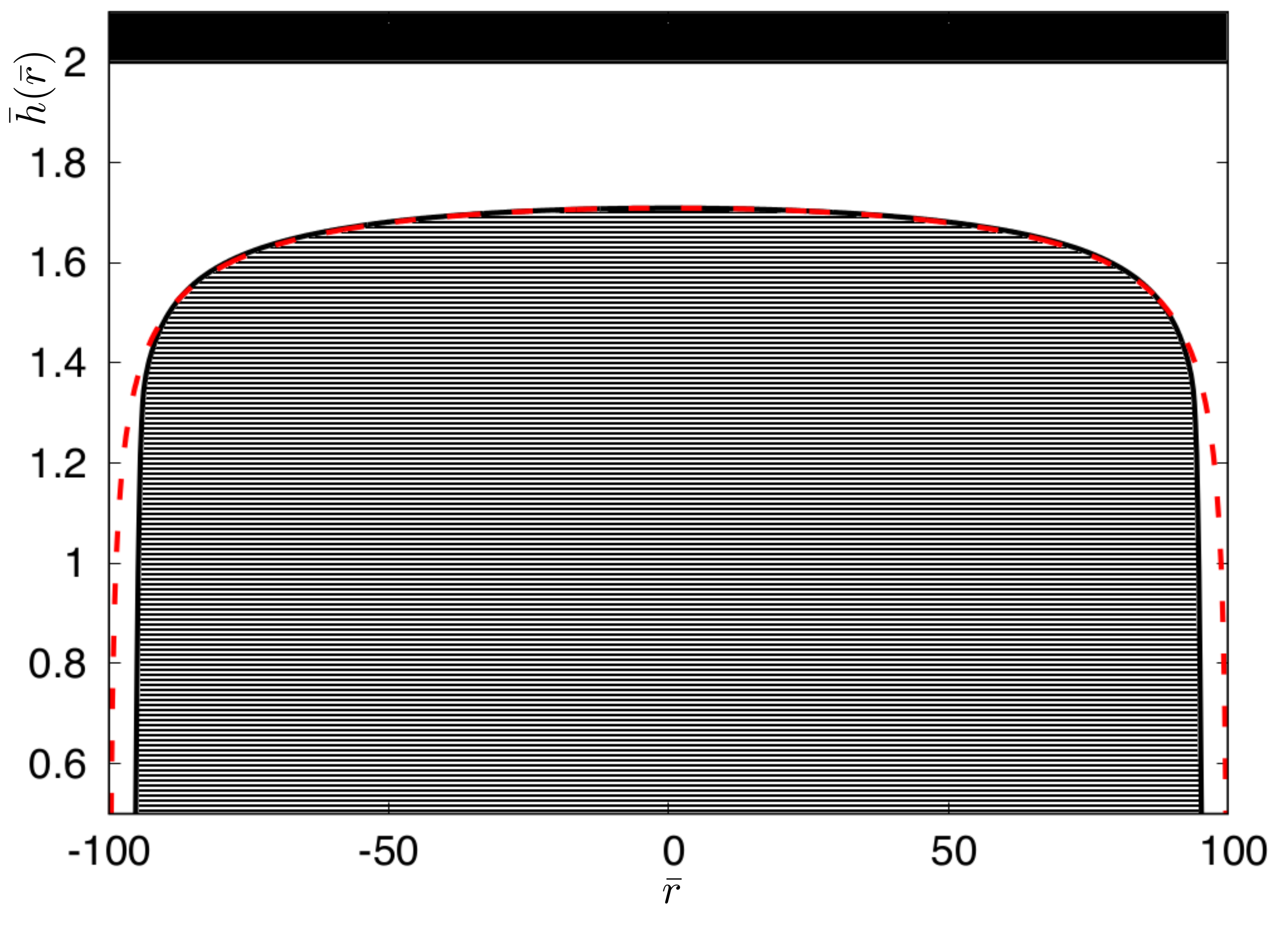}
\caption{Flattened steady-state for power-law repulsion.
Cross section of the steady state profile projected along $\bar{r}$ (solid line) dissolving under an external load, $\bar{F}_{Cz} = 10^8$, at $\bar{\eta} = 1$ against a flat substrate ($\bar{h}_s=2$).
Geometry: axisymmetric contact in a simulation box of size $\bar{R}=100$.
The interaction with the substrate is a singular power law repulsion, \cref{eq:singular_pot} with $n=3$. 
The red dashed line is the analytical prediction from \cref{eq:profile_powerLaw} with $r_m=R$ and $\zeta_0 \approx 0.29$ as a fitting parameter.}
\label{fig:profile_singular_rep}
\end{figure}

Let us start with the analysis of the results for a singular power 
law repulsion between the crystal surface and the substrate. 
Combining \cref{eq:singular_pot} and \cref{eq:Utilde_definition}
we find
\begin{equation}
\label{eq:tilde_singular_pot}
\tilde{U}(\zeta) = \frac{-(n+1)A}{\zeta^{n}}.
\end{equation}
Inserting this expression in \cref{eq:general_largeForce_1d,eq:general_largeForce_2d}, 
provides us with the steady-state profile:
\begin{subequations}
\label{eq:profile_powerLaw}
\begin{align}
\zeta(x) = \left( \frac{\zeta_0^n}{1-{x^2}/{x^2_m}}\right)^{1/n},
\\
\zeta(r) = \left( \frac{\zeta_0^n}{1-{r^2}/{r^2_m}}\right)^{1/n}.
\end{align}
\end{subequations}
These profiles diverge at $x=x_m$ or $r=r_m$, which is related to the minimum
distance in the contact via \cref{eq:general_largeForce_contact_1d,eq:general_largeForce_contact_2d}
\begin{subequations}
\label{eq:div_zeta0}
\begin{align}
\label{eq:xm_zeta0}
x^2_m = \frac{2D_{e}(n+1) A }{\zeta_0^n u_{Cz}}\, ,
\\
\label{eq:rm_zeta0}
r^2_m = \frac{4D_{e}(n+1) A }{\zeta_0^n u_{Cz}}\, .
\end{align}
\end{subequations}
The distance  $x_m$ or $r_m$ at which the profile diverges
should a priori be distinguished from the size of the contact region.
Indeed far away from the substrate, the influence of the potential vanishes,
 and as a consequence surface-tension effects
should become  dominant, 
so that \cref{eq:general_largeForce} is not valid anymore.
Let us define $L_c$ as the half-width of the contact region in 1D,
and $R_c$ as the radius of the contact region in 2D.
An intuitive definition of the contact region is the
zone which is close enough to the substrate to be under
the influence of the interaction potential $U$.

For large contacts, we expect that the distance separating $x_m$ and $L_c$, or $r_m$ and $R_c$
should be negligible as compared to the size of the contact region.
As a consequence, we assume $x_m\approx L_c$ or $r_m\approx R_c$.
Furthermore, we perform simulations with a fixed  $\zeta_{bc}$,
which is large as compared to $\zeta_0$ but small as compared to 
the size $L$, or $R$ of the simulation box. Thus,
the contact region should fill most of the simulation box,
and finally we expect $x_m\approx L_c\approx L$ or
$r_m\approx R_c\approx R$.
In \cref{fig:profile_singular_rep} we show the steady 
state cross section 
obtained from the simulation (solid line) at large times, 
which is in good agreement with 
\cref{eq:profile_powerLaw} using $r_m = R$ (dashed line) and $\zeta_0$ as a fitting parameter.
Using \cref{eq:div_zeta0} and the fitted value of $\zeta_0$ we obtain
a value for $u_{Cz}$. For instance in 2D with $\bar{F} = 10^8$ and $\bar{R}=100$, 
this procedure leads to $\bar{\zeta}_0 = 0.290$ 
and $\bar{u}_{Cz} =0.022$ to be compared with
 $\bar{\zeta}_0 =0.291$ and $\bar{u}_{Cz} =0.016$ 
measured directly 
in the numerical solution of the full model. 
The agreement with the numerical results improves as the external load is increased.

Similar agreement is obtained in 1D. 
As a consequence, the profile is well predicted
at large forces, and we can safely use it in the force balance equation.

In 1D, using \cref{eq:profile_powerLaw} with \cref{eq:xm_zeta0} and $x_m = L_c$ we obtain from force balance \cref{eq:vel_1d}
\begin{equation}
\begin{split}
&\frac{F^{1D}_{Cz}}{L_c} = 24\eta\phi\Bigl(\frac{n+3}{n}\Bigr) \frac{n\sqrt{\pi}}{(n+3)}\Bigl(	\frac{1}{D_{e}A(n+1)} \Bigr)^{\frac{3}{n}}\Bigl(\frac{L_c^2}{2} u_{Cz}\Bigr)^{\frac{n+3}{n}} \\
&+\phi\Bigl(\frac{n+1}{n}\Bigr) 2n\sqrt{\pi}A^{-\frac{1}{n}}\Bigl(	\frac{1}{D_{e}(n+1)} \Bigr)^{\frac{n+1}{n}}\Bigl(\frac{L_c^2}{2} u_{Cz}\Bigr)^{\frac{n+1}{n}}\mkern-14mu ,
\label{eq:F_vs_v_singularRep:1d}
\end{split}
\raisetag{\baselineskip}
\end{equation}
where
\[
\phi(z) =  \frac{\Gamma(1+z)}{2\Gamma(\frac{3}{2}+z)} \, ,
\]
with $\Gamma$ the Euler-Gamma function. 

Similarly, in 2D force balance \cref{eq:vel_2d} imposes
\begin{equation}
\begin{split}
\label{eq:F_vs_v_singularRep:2d}
&\frac{F^{2D}_{Cz}}{\pi R_c^2} = 12\eta \frac{n^2}{(2n+3)(n+3)}\Bigl(\frac{1}{D_{e}A(n+1)} \Bigr)^{\frac{3}{n}}\Bigl(\frac{R_c^2}{4} u_{Cz}\Bigr)^{\frac{n+3}{n}} \\
&+ \frac{n^2}{2n+1}A^{-\frac{1}{n}}\Bigl(	\frac{1}{D_{e}(n+1)} \Bigr)^{\frac{n+1}{n}}\Bigl(\frac{R_c^2}{4} u_{Cz}\Bigr)^{\frac{n+1}{n}} ,
\end{split}
\raisetag{\baselineskip}
\end{equation}
(some technical details about the derivation of this relation
can be found in \cref{appendix:noSurf_pwLaw}).
Using \ref{eq:F_vs_v_singularRep:2d}, 
we find two
separate regimes depending on the value of $\eta$:
For large viscosities we identify a \emph{hydrodynamic regime}
\begin{subequations}
\begin{align}
u^{1D}_{Cz}&= C^{1D}_h L_c^{-\frac{3n +6}{n+3}}\Bigl(\frac{F_{Cz}^{1D}}{\eta}\Bigr)^{\frac{n}{n+3}} ,\label{eq:hydro_1d}\\
u^{2D}_{Cz}&= C^{2D}_h R_c^{-\frac{4n +6}{n+3}}\Bigl(\frac{F_{Cz}^{2D}}{\eta}\Bigr)^{\frac{n}{n+3}} ,\label{eq:hydro_2d}
\end{align}
\end{subequations}
while for small viscosities a \emph{diffusion regime} is found, with
\begin{subequations}
\begin{align}
u^{1D}_{Cz}&= C^{1D}_d L_c^{-\frac{3n +2}{n+1}}(F_{Cz}^{1D})^{\frac{n}{n+1}}\, ,\label{eq:diff_1d}\\
u^{2D}_{Cz}&=C^{2D}_{d} R_c^{-\frac{4n +2}{n+1}}(F_{Cz}^{2D})^{\frac{n}{n+1}}\, \label{eq:diff_2d}.
\end{align}
\end{subequations}
The expressions of the constants $C^{1D}_h, C^{2D}_h,C^{1D}_d,C^{2D}_{d}$ 
are reported in \cref{appendix:noSurf_pwLaw}.
In \cref{fig:vel_force_powerLaw_2d} we compare 
the prediction \cref{eq:hydro_2d,eq:diff_2d} using $R_c=R$
(solid and dashed lines) and the results in 2D obtained from 
the complete numerical solution of the model (circles). 
The analytical prediction is in good agreement 
with the numerical solution for large external loads. 

In order to probe the sensitivity of the results
with respect to the value of the film
thickness at the boundary  $\bar{\zeta}_{bc}$,
we monitored the consequences of the variation of $\bar\zeta_{bc}$. 
We found small quantitative effects but 
no influence on the qualitative behavior of the relevant observables.
This is exemplified 
with the variations of the dissolution rates in the 
top panel of \cref{fig:vel_force_powerLaw_2d}.

Using \cref{eq:div_zeta0} to eliminate $u_{Cz}$
in the expression of the force~\cref{eq:F_vs_v_singularRep:1d,eq:F_vs_v_singularRep:2d}, 
a relation between external load 
and the minimum thickness $\zeta_0$ can be obtained, 
which is found to be in good agreement with the simulations. 
For the sake of concision, 
the expression of this relation in 2D and its comparison with
the numerical solution of the full model are shown
in  Appendix C ( \cref{eq:F_vs_zita_singularRep:2d}, and
 \cref{fig:zita_force_powerLaw2d}).

As an additional remark \cref{eq:F_vs_v_singularRep:2d,eq:F_vs_v_singularRep:1d} 
show that there is no substantial difference between 
one and two dimensions except, 
as expected from dimensional analysis, 
a different scaling with the contact size. 
\begin{figure}
\includegraphics[width=\linewidth]{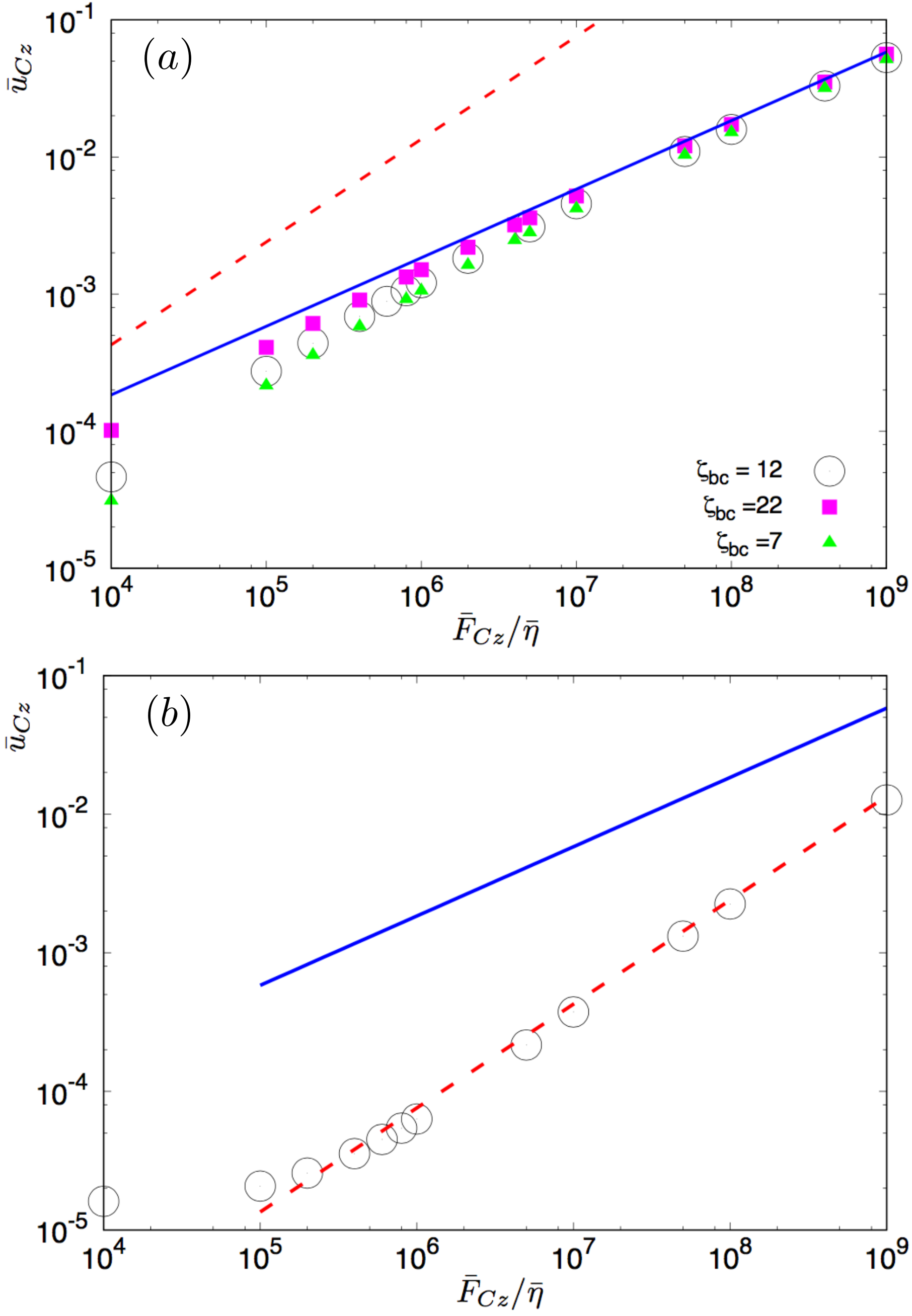}
\caption{Dissolution rate for (singular) power-law repulsion ($n=3$) as a function of the applied load.
Geometry: axisymmetric contact.  
Lines represent asymptotic analytical predictions with $\bar{R}_c = \bar{R} =100$: 
Solid blue line, hydrodynamic regime \cref{eq:hydro_2d};  
dashed red line, diffusion regime \cref{eq:diff_2d}.
(a): $\bar{\eta} =1$,
the symbols show simulation results obtained using different boundary thickness $\zeta_{bc}$.
(b): $\bar{\eta} = 10^{-3}$. The results are shown in normalized units.
}
\label{fig:vel_force_powerLaw_2d}
\end{figure}
Finally, it is interesting to assess what 
is the critical length and load separating the diffusive
and hydrodynamic regimes.
Equating the expressions of the velocity
in the two regimes for the ridge case, 
we find  that the critical size above which 
the force is dominated by the diffusion term is given by
\begin{equation}
\label{eq:critical_L_1d}
L_*= B_{1D}
\eta^{\frac{n+1}{2}}F_{Cz}^{1D} \, ,
\end{equation}
in 1D, while for the axisymmetric contact is
\begin{equation}
\label{eq:critical_L_2d}
R_*= B_{2D}\eta^{\frac{n+1}{4}}(F_{Cz}^{2D})^{1/2} \, ,
\end{equation}
where $B_{1D}$ and $B_{2D}$ are constants reported in \cref{appendix:noSurf_pwLaw}. 
Hence,  at fixed force large contacts will be dominated by the diffusion term.
Also, as the external load is increased at constant contact size 
the hydrodynamic term in the force balance become dominant. 
Once again, good agreement with the simulations is found,
and a detailed discussion is reported in \cref{appendix:noSurf_pwLaw}.

\subsection{Finite repulsion: exponential case }
\label{sec:exp_rep}

\begin{figure}
\includegraphics[width=\linewidth]{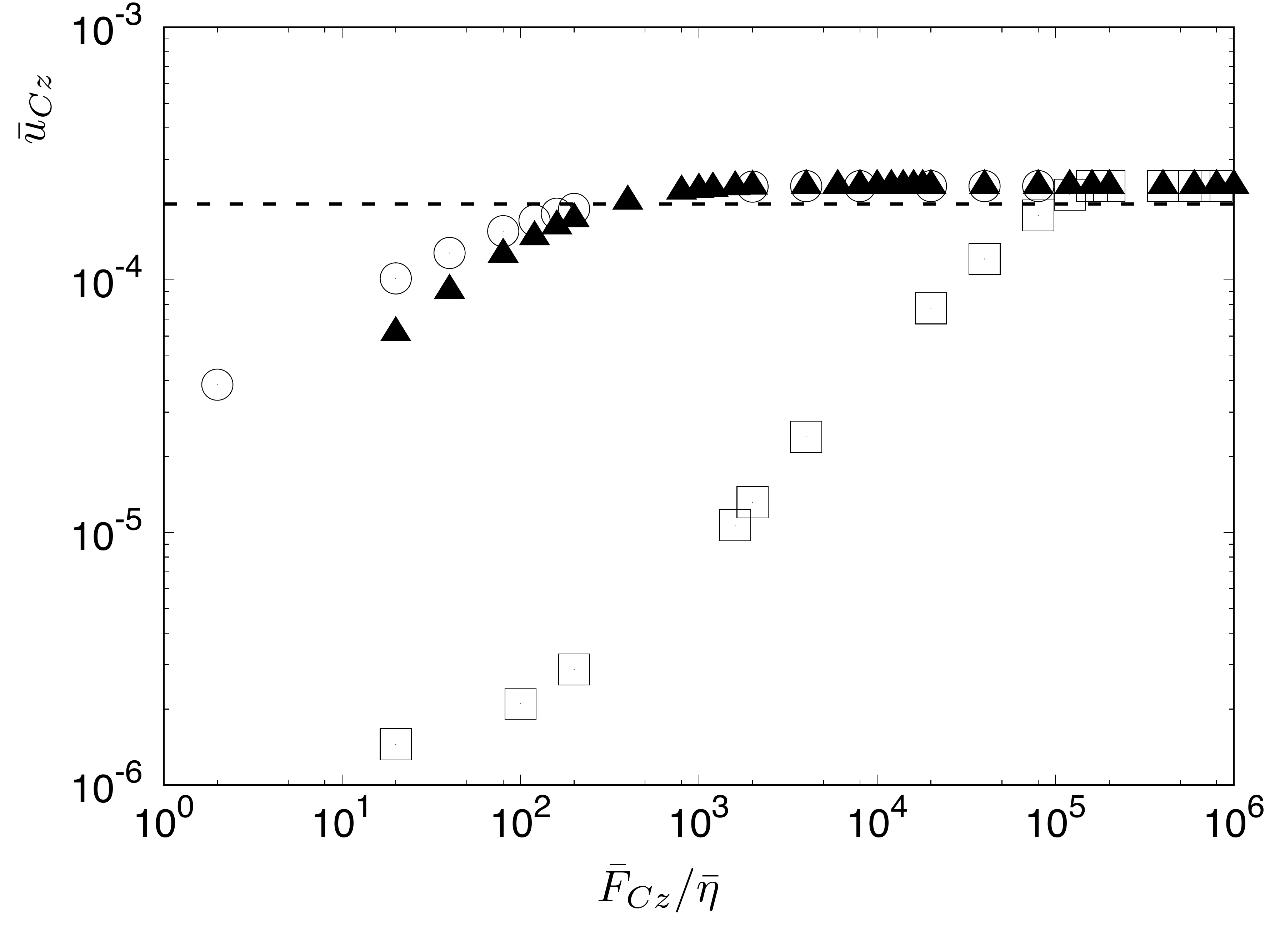}
\caption{Dissolution rate as a function of the external load for an exponential finite interaction.
Geometry: 1D ridge contact. 
Triangles $\bar{\eta} =0.5$, squares $\bar{\eta} =5\times 10^{-4}$, 
circles $\bar{\eta} =5\times 10^2$. 
Dashed line: analytical prediction \cref{eq:vel_exp_1d} using $\bar{L}_c = \bar{L}=100$ and $\zeta_0/\lambda = 0$. 
The results are in normalized units.}
\label{fig:exp_vel}
\end{figure}

In the case of an exponential repulsion, $\tilde{U}(\zeta=0)$ is finite. 
As a consequence, the behavior of steady-state solutions is different. 
First, the dissolution rate is asymptotically independent of the load. 
Second, the shape of the contact is sharp and pointy.
Third, in the absence of surface tension,
touching contact (i.e. $\zeta=0$)  would be observed in 2D
for a finite loading force, but not in 1D. 
Finally, as opposed to what observed so far, 
surface tension becomes relevant at large enough forces 
both in 1D and 2D, and prevents contact also in 2D.

\subsubsection{Without surface tension}
\label{sec:exp_noSurf}
Neglecting surface tension we  proceed in a similar way as in the power-law case.  
Recalling \cref{eq:Utilde_definition} and using the
exponential interaction potential \cref{eq:interaction_exp},
we now have
\begin{equation}
\label{eq:exp_rep}
\tilde{U}(\zeta) = -\frac{A}{\lambda}(\lambda + \zeta)e^{-\frac{\zeta}{\lambda}}\, .
\end{equation}
As opposed to the power-law repulsion case,
we now have a function $\tilde{U}(\zeta)$
that cannot be inverted explicitly.
Therefore, $\zeta$ cannot be explicitly obtained
from \cref{eq:general_largeForce}. However, since $\tilde{U}$ is a monotonic
function of $\zeta$, 
it is still possible to compute $r$ as a function of $\zeta$
without ambiguity from \cref{eq:general_largeForce}.

In the large force limit since we expect $\zeta_0\ll\lambda$ (this will be confirmed below
using force balance) and since $\tilde U(0)$ is finite, 
we find that the dissolution rate reaches a constant value independent of the load and of the viscosity. 
Indeed, from \cref{eq:general_largeForce_contact}:
\begin{subequations}
\label{eq:vel_exp_zeta0}
\begin{align}
\label{eq:vel_exp_1d_zeta0}
u_{Cz}^{1D} \approx D_e\frac{2A}{L_c^2}(1+\frac{\zeta_0}{\lambda})e^{-\frac{\zeta_0}{\lambda}} ,\\
\label{eq:vel_exp_2d_zeta0}
u_{Cz}^{2D} \approx D_e\frac{4A}{R_c^2}(1+\frac{\zeta_0}{\lambda})e^{-\frac{\zeta_0}{\lambda}} .
\end{align}
\end{subequations}
Taking the limit $\zeta_0\rightarrow 0$, we find
\begin{subequations}
\label{eq:vel_exp}
\begin{align}
\label{eq:vel_exp_1d}
u_{Cz}^{1D}=D_{e}\frac{2A}{L_c^2} \, ,\\
\label{eq:vel_exp_2d}
u_{Cz}^{2D} = D_{e}\frac{4A}{R_c^2}\, .
\end{align}
\end{subequations}
Again assuming that $L_c\approx L$,
or $R_c\approx R$ at large forces, 
these results are confirmed in \cref{fig:exp_vel}
from the comparison with the numerical solution of the full model. 
The different viscosities, indicated by circles ($\bar{\eta}=1000$), 
triangles ($\bar{\eta} = 1$) and squares ($\bar{\eta} = 0.001$), 
affect the absolute value of the applied force needed to reach the plateau but not the plateau value itself.

\begin{figure}
\includegraphics[width=\linewidth]{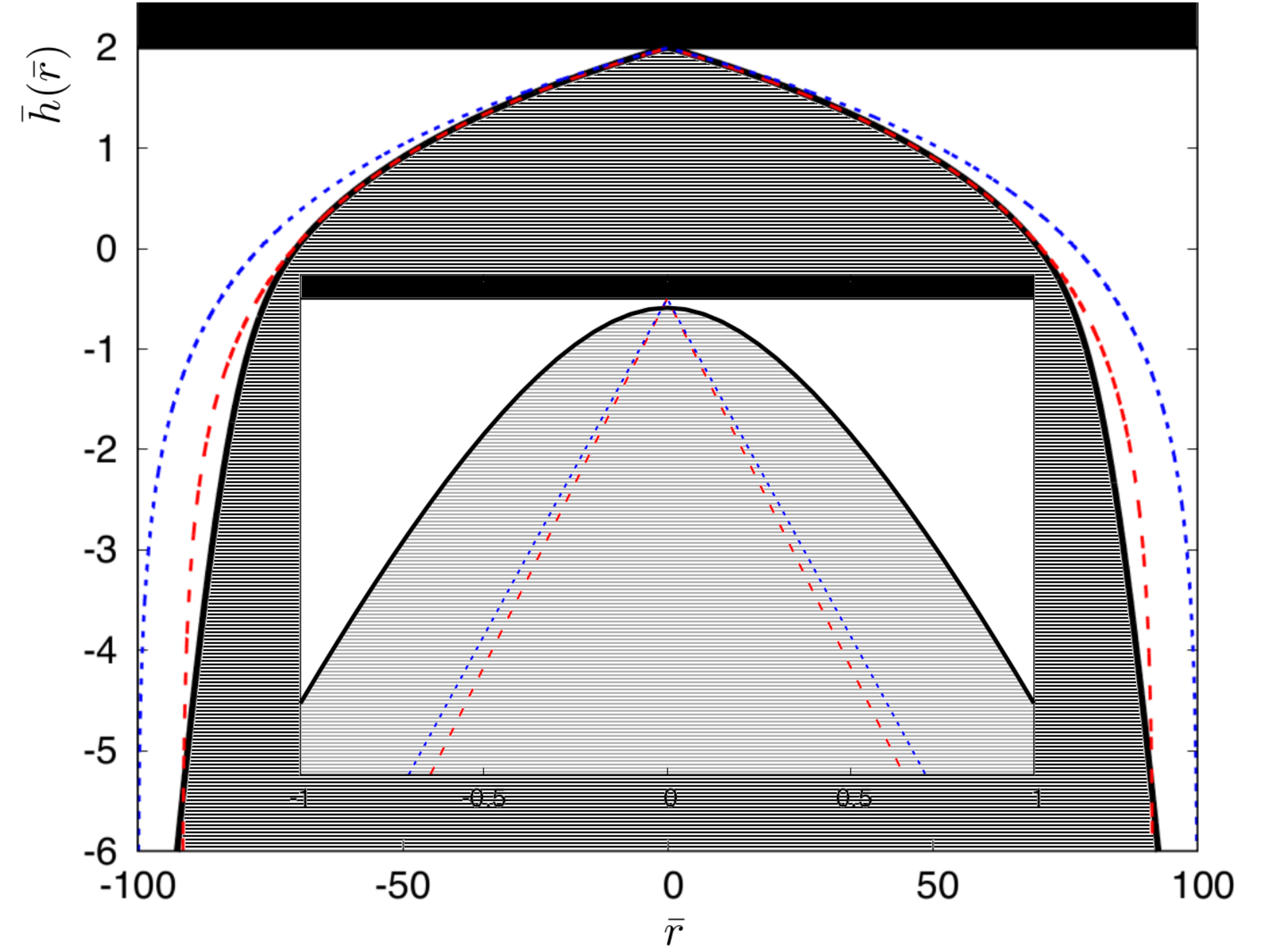}
\caption{Pointy steady-state for exponential repulsion.
Cross section of the steady state profile projected along $\bar{r}$ (solid line) 
dissolving under an external load $\bar{F}_{Cz} = 1.7\,\, 10^5$, at $\bar{\eta} = 1$ 
against a flat substrate ($\bar{h}_s=2$).
Geometry: axisymmetric contact in a simulation box of size $\bar{R}=100$.
The interaction with the substrate is a finite exponential repulsion, \cref{eq:interaction_exp}. 
Blue dotted line:  analytical prediction \cref{eq:height_2d} 
assuming the contact area to be equal to the surface size $R$.
Red dashed line:  analytical prediction \cref{eq:height_2d} 
with a smaller contact size $R_c$. 
The inner plot shows a zoom of the tip.}\label{fig:exp_shape}
\end{figure}

A second consequence arising from the finiteness of the
exponential interaction 
is the sharp pointy shape of the steady state profile 
showed in \cref{fig:exp_shape}. 
Indeed, since  $\tilde U'(\zeta=0)=0$ from \cref{eq:Utilde_definition},
we have   $\tilde U(\zeta)\approx\tilde U(0)+\tilde U''(0)\zeta^2/2$
for $\zeta\ll\lambda$. 
Using this expansion into \cref{eq:general_largeForce} and letting
$\zeta_0\rightarrow 0$, 
we find that the profile $\zeta_{sing}$ 
in the center of the contact region 
is a singular wedge in 1D and a cone in 2D :
\begin{subequations}
\label{eq:pointy_shape}
\begin{align}
\label{eq:pointy_shape_1d}
\zeta_{sing}\approx  
\left(\frac{u_{Cz}}{D_{e} \tilde U''(0)}\right)^{1/2} |x|
=\left(\frac{u_{Cz}}{D_{e} A}\right)^{1/2} \lambda |x|\, ,\\
\label{eq:pointy_shape_2d}
\zeta_{sing}\approx  
\left(\frac{u_{Cz}}{2 D_{e} \tilde U''(0)}\right)^{1/2} |r|
=\left(\frac{u_{Cz}}{2 D_{e} A}\right)^{1/2} \lambda |r|\, .
\end{align}
\end{subequations}
When $\zeta_0\ll\lambda$,
the complete profile for arbitrary $\zeta$
(i.e. smaller or larger than $\lambda$) can be obtained from \cref{eq:exp_rep,eq:general_largeForce}.
Using the axisymmetric contact, with $R_c = R$ and $u_{Cz}$ 
given by \cref{eq:vel_exp_2d} this expression (dotted blue line) is 
seen to be in good agreement with the simulation in \cref{fig:exp_shape}. 
Better agreement (red dashed line) can be reached using the numerical value 
of $u_{Cz}$ obtained from the simulation (which is equivalent to assuming a smaller effective size, $R_c<R$). 
Nevertheless as showed by the inner panel in  \cref{fig:exp_shape}, 
close to the tip the numerical solution is smoothed and exhibits a parabolic
shape. This regularization  discussed in the next section 
is due to the contribution of the surface tension.

Using \cref{eq:vel_1d}, force balance in 1D now reads:
\begin{equation}
\label{eq:force_exp1d}
\frac{F^{1D}_{Cz}}{L_c} =\Bigl [ 12\eta D_{e}\frac{A}{\lambda^3}\psi_1(\frac{\zeta_0}{\lambda}) + \frac{A}{\lambda} \psi_2(\frac{\zeta_0}{\lambda}) \Bigr]\Bigl (\frac{e^{\frac{\zeta_0}{\lambda}}}{1+\frac{\zeta_0}{\lambda}}\Bigr)^{\frac{1}{2}}\, ,
\end{equation}
where the function $\psi_1$ and $\psi_2$ defined in \cref{eq:psi_1,eq:psi_2} exhibit the following limits
\begin{align*}
\lim_{z\to 0}\psi_1(z) &=\sqrt{2}\ln \frac{1}{z} + C_1 \, ,\\
\lim_{z\to 0}\psi_2(z)&= C_2 \, ,
\end{align*}
with $C_1 \approx 1.645$ and $C_2 \approx 0.8398$. 
It follows that, when $\zeta_0\ll \lambda$ and $\zeta_0\ll \lambda \exp[-C_2/(12\sqrt{2}\bar\eta)]$,
we have
\begin{equation}
\label{eq:cryticalF_1d}
F^{1D}_{Cz}\approx 12\sqrt{2} \bar\eta\frac{L_cA}{\lambda} \ln(\frac{\lambda}{\zeta_0}) \, .
\end{equation}
This relation indicates that the minimum distance in the contact region
decreases exponentially with the applied load in 1D.
The prediction \cref{eq:force_exp1d} using $L_c = L$, 
which is represented in \cref{fig:force_exp_comparison} 
by the red solid line, compares well with the numerical results (red circles)
when $\zeta_0$ is not too small.

In addition,  we obtain in 2D 
(some details of the derivation are reported 
in \cref{appendix:noSurf_exp})
\begin{equation}
\label{eq:force_exp2d}
\frac{F^{2D}_{Cz}}{\pi R_c^2} = \Bigl [ 12\eta D_{e}\frac{A}{\lambda^3}\psi(\frac{\zeta_0}{\lambda
})\frac{e^{\frac{\zeta_0}{\lambda}}}{1+\frac{\zeta_0}{\lambda}} +\frac{A}{4\lambda}(\frac{2\zeta_0}{\lambda}+1)\frac{e^{-\frac{\zeta_0}{\lambda}}}{1+\frac{\zeta_0}{\lambda}}\Bigr]\, ,
\end{equation}
where the function $\psi$
obeys
\[
\lim_{z\to 0} \psi(z) = (1-\ln 2)\, .
\]
Hence, within this approximation, the LC interface touches the substrate (i.e. $\zeta_0=0$) for a finite force 
\begin{equation}
\label{eq:cryticalF_2d}
F^{2D}_c = \Bigl [ 12\eta D_{e}\frac{A}{\lambda^3} (1-\ln(2) ) 
+ \frac{A}{4\lambda} \Bigr]\pi R_c^2 \, .
\end{equation}
The external force is plotted as a function of $\zeta_0$ in \cref{fig:force_exp_comparison}.
\Cref{eq:force_exp2d} with $R_c = R$ is represented by the blue solid line  and has to be compared with the blue squares obtained by direct numerical integration.
Once again, this expression agrees with the numerical
results for $\zeta_0$ large enough.

\subsubsection{With surface tension}

\begin{figure}
\includegraphics[width=\linewidth]{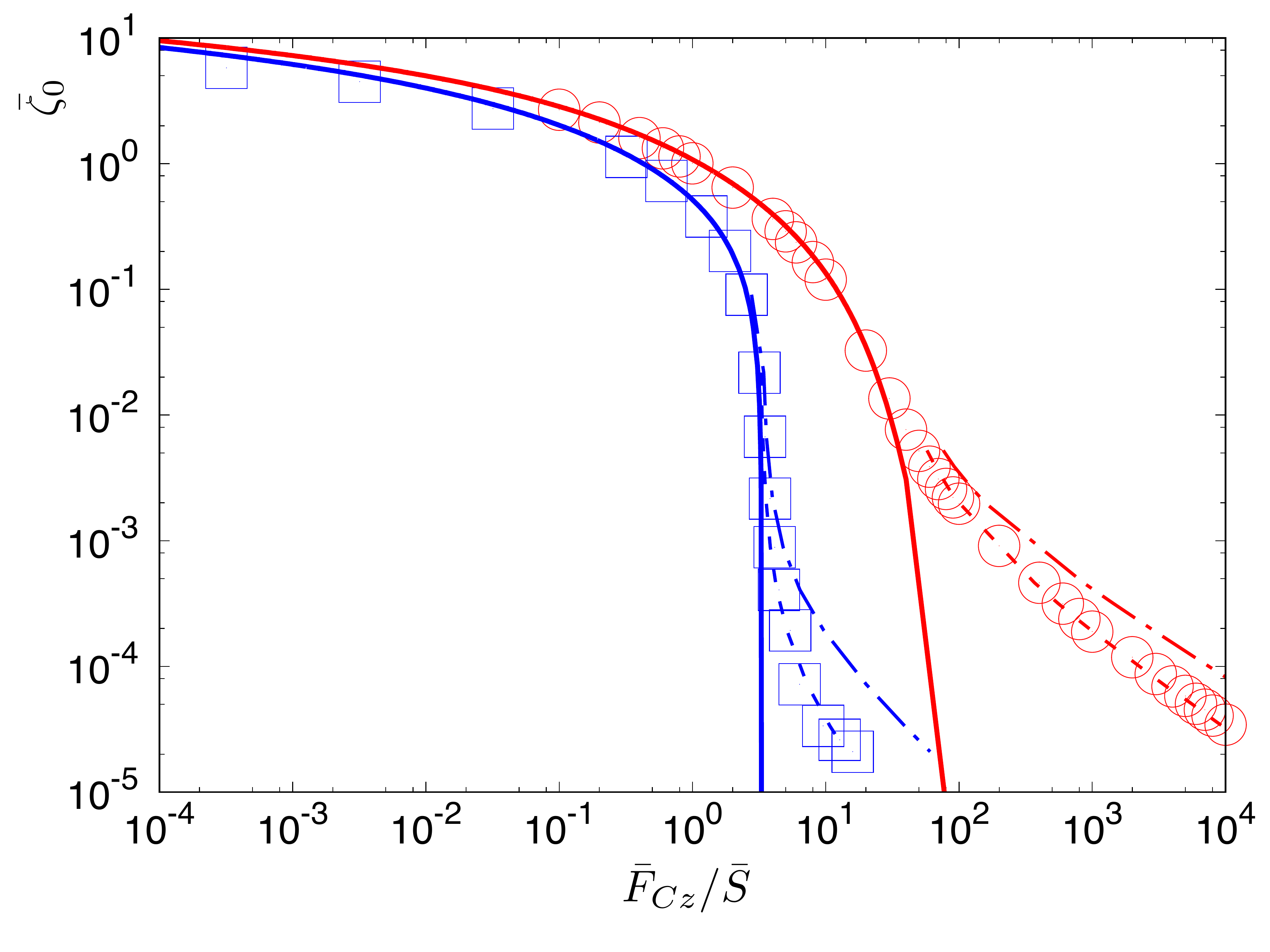}
\caption{ Minimum film thickness of the liquid film as a function of the applied load.
The plot shows the minimum distance $\bar{\zeta}_0$ between the crystal and the substrate
versus the external load normalized by surface area $\bar{S}$ (scaled pressure). 
Red, ridge contact (1D); blue, axisymmetric contact (2D). 
Circles (1D) and squares (2D) show the numerical results; 
Solid lines report the analytical predictions neglecting surface tension \cref{eq:force_exp1d,eq:force_exp2d} blue and using $\bar{L}_c=\bar{L}=100$, $\bar{R}_c=\bar{R}=100$; 
Dashed lines:  prediction  adding the singular contribution of the surface tension term \cref{eq:Ftip_complete_2d,eq:Ftip_complete_1d}
to the previous expression, and using the parameters $u_{Cz}$ and $\partial_{xx}\zeta_0$
or $\partial_{rr}\zeta_0$ from the simulations. 
Dashed-dotted lines: full analytical prediction using \cref{eq:F_tip_1d,eq:F_tip_2d}.
1D viscosity, $\bar{\eta} =0.5$; 2D viscosity, $\bar{\eta} =1$. 
The results are in normalized units. 
The critical force in 2D~\cref{eq:cryticalF_2d} provides the maximum value of $\bar F_{Cz}^{2D}/\bar{S}$ for the solid blue line 
and corresponds to $p \approx 3.3$MPa. 
}
\label{fig:force_exp_comparison}
\end{figure}

An inspection of \cref{fig:force_exp_comparison} reveals
that the agreement between the predicted force-minimum distance relation 
and the full numerical solution of thin film equations 
is accurate only when the forces are not too large. 
However, as we keep increasing the external load, 
this prediction (solid lines) fails to reproduce 
the numerical results. 
As anticipated previously, the shape of the crystal close to the tip
(see inner panel of \cref{fig:exp_shape}) is not well described by
\cref{eq:general_largeForce_contact}.
Indeed, as $\zeta_0\rightarrow 0$, the curvature
at the tip diverges, leading to the singular
pointy shape reported in \cref{eq:pointy_shape}. 
Thus,  surface tension
effects proportional to the curvature become relevant. 

We here resort to a simple matching procedure
to account for the consequences of surface tension.
First, in the tip region for $x<x_*$ or $r<r_*$, 
 where $x_*$ and $r_*$ are the tip width in 1D and 2D respectively,
a Taylor expansion of $\zeta$ leads to:
\begin{subequations}
\label{eq:tip_parabola}
\begin{align}
\label{eq:tip_parabola_1d}
\zeta^{tip} = \zeta_0 + \frac{x^2}{2}\partial_{xx}\zeta_0 \, ,\\
\label{eq:tip_parabola_2d}
\zeta^{tip} = \zeta_0 + \frac{r^2}{2}\partial_{rr}\zeta_0 \, ,
\end{align}
\end{subequations}
where $\partial_{rr}\zeta_0$ and  $\partial_{xx}\zeta_0$ 
are the second derivative of $\zeta$ calculated at $x=0$ or $r=0$.

Using this solution let us compute the  contribution 
of the tip region to force balance \cref{eq:vel_1d,eq:vel_2d}. We obtain
\begin{subequations}
\begin{align}
\label{eq:Ftip_complete_1d}
F^{1D}_{\text{tip}}=& 2\frac{Ax_*}{\lambda}
\Bigl (1-\frac{\zeta_0}{\lambda} 
- \partial_{xx}\zeta_0\frac{x_*^2}{6\lambda}  \Bigr ) 
+\eta\frac{6\pi u_{Cz}}{\sqrt{2}(\partial_{xx}\zeta_0)^{3/2}\zeta_0^{3/2}} \, ,
\\
\label{eq:Ftip_complete_2d}
F^{2D}_{\text{tip}}= &  \frac{\pi Ar_*^2}{\lambda} 
\Bigl(1 - \frac{\zeta_0}{\lambda} 
- \partial_{rr}\zeta_0 \frac{r_*^2}{4\lambda}\Bigr) 
+ \eta\frac{6\pi u_{Cz}}{(\partial_{rr}\zeta_0)^2\zeta_0} \, ,
\end{align}
\end{subequations}
where we used $\zeta/\lambda \ll 1$ in the tip region.
From this expression it appears that, if $x_*$ or $r_*$
is not increasing too fast when the load increases 
and $\zeta_0\rightarrow 0$,
the dominant term is the one proportional to the viscosity. 

To confirm the validity of this statement, we checked that
the increase of the force at small $\zeta_0$
is well predicted by adding  the singular contribution 
corresponding to the last term of
\cref{eq:Ftip_complete_1d,eq:Ftip_complete_2d}  
to the previous expressions. The result reported in \cref{fig:force_exp_comparison},
agrees well with the deviations at small $\zeta_0$. 
However, this relation is still not fully predictive, 
since we used $\partial_{xx} \zeta_0$ and $\partial_{rr} \zeta_0$ obtained 
from the numerical solution. 
In order to find an additional relation linking $\zeta_0$ 
and $\partial_{xx} \zeta_0$ or $\partial_{rr}\zeta_0$, we match the solutions 
far from and close to the tip in the limit $\zeta \ll \lambda$. 

Far from the tip, we assume a small deviation $\delta\zeta$
from the singular solution \cref{eq:pointy_shape},
leading to $\zeta=\zeta_{sing}+\delta\zeta$.
To find an expression for $\delta \zeta$
we insert the previous relation into the full steady-state differential equation in the presence of curvature terms 
\begin{subequations}
\begin{align}
\label{eq:steady_1d}
0 & = u_{Cz} + D_{e}
\partial_x\Bigl[\zeta \Bigl(
\gamma \partial_{xx}\zeta - U'(\zeta)
\Bigr )\Bigr ]\, ,\\
\label{eq:steady_2d}
0 & = u_{Cz} 
+ D_{e}\frac{1}{r}\partial_r\Bigl[r\zeta \Bigl(\gamma \partial_{rr}\zeta 
+\frac{\gamma}{r}\partial_r \zeta - U'(\zeta)\Bigr )\Bigr ]\, ,
\end{align}
\end{subequations}
for the 1D and 2D respectively.
Matching the height and the slope
of the tip solution \cref{eq:tip_parabola}
with the perturbative solution outside the tip
region $\zeta=\zeta_{sing}+\delta\zeta$
at some position $x_*$ or $r_*$
leads to two equations. These
two equations are used to obtain
$x_*$ or $r_*$, and $\partial_{xx}\zeta_0$
or $\partial_{rr}\zeta_0$,
as a function of $\zeta_0$.
We therefore have a profile with two regions
that is completely determined by $\zeta_0$.
The details of the derivations is quite cumbersome,
and is therefore reported
in \cref{appendix:derivation_expTip}.

Two important remarks are in order.
First, due to the correction $\delta\zeta$, 
the profile becomes wider when approaching 
the tip region in agreement with the shape observed in the full numerical solution in \cref{fig:exp_shape}.

As a second remark, the matching analysis 
shows that $\partial_{xx}\zeta_0$ and $\partial_{rr}\zeta_0$ 
tend to a constant for  $\zeta_0\rightarrow 0$.  
Using these results in the expression of the force,  
we obtain asymptotically a power law dependence of the force  on $\zeta_0$
\begin{subequations}
\begin{align}
\label{eq:F_tip_1d}
F^{1D}_{\text{tip}} & = \frac{12\pi\eta \gamma^{3/2}D_{e}\lambda^{3/2}}{C_{1D}^{3/2}\sqrt{2A}L_c^2} \frac{1}{\zeta_0^{3/2}}
+ \text{ non singular terms} \, ,\\
\label{eq:F_tip_2d}
F^{2D}_{\text{tip}} & = \frac{24\pi\eta\gamma^2 D_{e}\lambda^2}{C_{2D}^2 A R_c^2}\frac{1}{\zeta_0} + \text{ non singular terms}\, ,
\end{align}
\end{subequations}
 where the constants $C_{1D}=\partial_{\bar{x}\bar{x}}\bar{\zeta}_0(\bar{\zeta_0}= 0)$ 
and $C_{2D}=\partial_{\bar{r}\bar{r}}\bar{\zeta}_0(\bar{\zeta_0}= 0)$
are the values of the normalized second derivatives at the tip when
$\bar\zeta_0\rightarrow 0$. 
From simulations, we find $C_{1D} \approx 0.017$ and $C_{2D}\approx 0.015$
(see \cref{fig:curv_vs_zita}).
Note that we used the approximated expression of 
the dissolution rates $u_{Cz}$ given by \cref{eq:vel_exp}.

The sum of the contribution without surface tension \cref{eq:force_exp1d,eq:force_exp2d} 
with the contribution of the tip  \cref{eq:F_tip_1d,eq:F_tip_2d}, 
are presented in \cref{fig:force_exp_comparison} by the dashed-dotted lines. 
The agreement with the full numerical solution is not quantitative, 
but is satisfactory considering the heuristic character of the matching procedure.
Fitting the numerical results with power laws at large forces,  
we obtain for the wedge-like  contact  $F_{Cz}\sim \zeta_0^{-1.3}$ 
to be compared 
with the prediction  $F_{Cz}\sim \zeta_0^{-3/2}$ from \cref{eq:F_tip_1d}, 
while for the conical contact $F_{Cz}\sim \zeta_0^{-1.1}$ 
to be compared with $F_{Cz}\sim \zeta_0^{-1}$ from \cref{eq:F_tip_2d}. 

As a final comment, the critical force for which 
surface tension becomes relevant
is given by \cref{eq:cryticalF_2d} in 2D.
In 1D, comparing \cref{eq:cryticalF_1d,eq:F_tip_1d} by 
\[
F^{1D}_{c} \approx 24\eta D_e\frac{A}{2\lambda^3} L_c\, ,
\]
up to logarithmic corrections.

\section{Discussion }
\label{sec:discussion}

\subsection{Summary of results}
\label{sec:summary}

In this paper, we have obtained a thin film model describing 
the evolution of  a rigid crystal that is able to grow or dissolve,
in the vicinity of a substrate. 
The model includes hydrodynamics, diffusion, 
the disjoining pressure effects, and surface tension.

Using this model, we have studied pressure solution 
against a flat wall in ridge-like (1D), 
and axisymmetric (2D) contacts.
This study has been performed using some simplifying assumptions, 
including equal-density between the liquid and the crystal, the 
linearization of Gibbs-Thomson relation and the dilute approximation.

We have also considered two different types of repulsions 
between the substrate and the crystal. 
These led to different  behaviors.

In the case of a power-law repulsion diverging at contact, 
the crystal interface flattens under load, 
and the dissolution rate exhibits a power-law dependence on the load.
A change in this power-law is found at large loads and viscosities
when the forces induced by viscous dissipation 
surpass those due to disjoining pressure.

In contrast,  a finite exponential repulsion produces 
pointy contacts and a dissolution rate 
asymptotically independent of the load and of the viscosity.
For large loads, the sharp pointy shape of the tip 
is regularized by surface tension,
and the force balance is dominated by viscous effects.
Touching contact (i.e. $\zeta_0=0$) is found only in 2D and in the absence of surface tension.

To summarize, we found that for large external loads the dissolution rate $u_{Cz}$ and minimum distance $\zeta_0$ between the dissolving crystal 
and the substrate obey scaling laws
\begin{subequations}
\label{eq:summary}
\begin{align}
\label{eq:summary_1d}
u_{Cz}&\sim F^{\alpha_u}_{Cz}L_c^{\beta_u}\quad
\zeta_0\sim F^{\alpha_\zeta}_{Cz} L_c^{\beta_\zeta}\\
\label{eq:summary_2d}
u_{Cz}&\sim F^{\alpha_u}_{Cz}R_c^{\beta_u}\quad
\zeta_0\sim F^{\alpha_\zeta}_{Cz} R_c^{\beta_\zeta} \, ,
\end{align}
\end{subequations}
where $F_{Cz}$ is the external load and $L_c$ or $R_c$ are the contact sizes for the ridge and the axisymmetric contact, respectively. 
The exponents $\alpha_u$, $\beta_u$, $\alpha_\zeta$, $\beta_\zeta$ 
displayed in \cref{tab:exponents} are found to depend 
on dimensionality (ridge or axisymmetric), on viscosity, 
and on the type of interaction potential 
(diverging as a power-law or finite at contact).

\begin{table*}
\caption{Summary of the asymptotic regimes
at large loads following the notation of \cref{eq:summary}. 
For exponential potentials in the absence of surface tension effects,
the dependence of $u_{Cz}$ and $\zeta_0$ on the load and system size is not a power-law.
In 1D the dependence is logarithmic \cref{eq:cryticalF_1d}, and in 2D $\zeta_0$  vanishes (i.e. the crystal touches the substrate)
for a finite force $F_c^{2D}$ \cref{eq:cryticalF_2d}. 
}
\label{tab:exponents}
\renewcommand{\arraystretch}{1.3}
\begin{tabular}{|c|c|c|c|c|c|c|}
\hline
\textrm{  Repulsion } &
\multicolumn{2}{c|}{Power-Law } &
\multicolumn{4}{c|}{Exponential } \\
\hline
\textrm{  Regime  } &
\textrm{ Hydrodyn. }&
\textrm{ Diffusion }&
 \textrm{ 1D no surf.\ tens.}&
 \textrm{ 1D surf.\ tens.}&
\textrm{ 2D no surf.\ tens.}&
\textrm{ 2D surf.\ tens.}\\
\textrm{ } &
\textrm{$\bar \eta \geq 1$ }&
\textrm{$\bar \eta \ll 1$ }&
 \textrm{ $F_{Cz}^{1D}\ll F^{1D}_c$}&
 \textrm{$F_{Cz}^{1D}\gg F^{1D}_c$ }&
\textrm{$F_{Cz}^{2D}<F^{2D}_c$}&
\textrm{$F_{Cz}^{2D}\gg F^{2D}_c$}\\
\hline
$\alpha_u$	&	$\frac{n}{n+3}$	&	$\frac{n}{n+1}$	&	0(constant)	&	0(constant)&	0(constant)&	0(constant)\\
$\beta_u$ 	&	$-\frac{4n +6}{n+3}$	&	$-\frac{4n + 2}{n+1}$	&	$-2$	&	$-2$&	$-2$ &	$-2$\\
\hline
$\alpha_\zeta$	&	$-\frac{1}{n+3}$	&	$\frac{-1}{n+1}$	&	 exponential	& $-2/3$&
$\zeta_0\rightarrow 0$ as $F_{Cz}^{2D}\rightarrow F^{2D}_c$	&	$-1$\\
$\beta_\zeta$	&	$\frac{2}{n+3}$	&	$\frac{2}{n+1}$	&	 exponential	& $-4/3$	&	 &	$-2$\\
\hline
\end{tabular}
\end{table*}

\subsection{Orders of Magnitude and model limitations}
\label{sec:order_magnitude}

Before discussing precise systems,
we provide some orders of magnitude describing the energy
scale of the interactions. 
Various experiments and standard textbooks~\cite{Israelachvili1991} indicate that
the order of magnitude of disjoining pressures is typically
$U'\sim $MPa when the 
distance between the surfaces is $\zeta\sim \mathrm{nm}$.
For exponential interactions with decay length $\lambda\sim \mathrm{nm}$
(corresponding e.g.\ to the Debye length
or to hydration scales),
we obtain that $A\sim\lambda U'\sim \mathrm{mJ}\cdot\mathrm{m}^{-2}$.
As a consequence, the dimensionless repulsion strength (see \cref{sec:normalization}) is 
$\bar{A}=A/\gamma\sim 10^{-2}$.
For power-law interactions, with a typical
distance $\lambda\sim \mathrm{nm}$, we have
$A\sim U'\lambda^{n+1}$. As a consequence,
we also find $\bar{A}=A/\gamma \lambda^n\sim 10^{-2}$.

We now consider two different crystals: calcite CaCO$_3$, and sodium chlorate NaClO$_3$.
For calcite we use\cite{Plummer1976,Li1973}: 
solubility $c_0 \approx10^{-3}\mathrm{mol/l}\approx10^{24}/\mathrm{m^3}$
(at 25$^{\circ}$C), 
molecular volume $\Omega \approx 100\AA^3$, 
ionic diffusion constant $D\approx 10^{-5}\mathrm{cm^2/s}$, 
water-solution interfacial tension\cite{Voort1988} $\gamma \approx 100\mathrm{mJ}$ 
and $T\approx 300$K.

For each variable  $y$ in physical units, 
and the corresponding variable $\bar{y}$ in 
normalized units, we define the scaling factor
$s_y$ from the relation $y= s_y\bar{y}$.
These scaling factors have to be applied 
to the simulation results to recover physical units.
Their precise expressions are given in \cref{appendix:scaling}.
In the case of Calcite, we estimate from \cref{eq:scalings}:
\begin{align*}
s_\zeta&=\mathcal{O}(1\,\mathrm{nm})\\
s_x &=\mathcal{O}(10\, \mathrm{nm})\\
s_t &=\mathcal{O}(10^{-1} \mathrm{s})\\ 
s_p &= \mathcal{O}(\mathrm{M\,Pa})\\ 
s_\eta &=\mathcal{O}(10^2 \mathrm{Pa\, s}) \, .
\end{align*}

Considering now NaClO$_3$ with \cite{Wang1996,Misbah2004}
$c_0\approx10^{28}/\mathrm{m^3}$(at 25$^{\circ}$C), $\Omega \approx 100\AA^3$, 
$D\approx 10^{-5}\mathrm{cm^2/s}$, $\gamma \approx 10\mathrm{mJ}$ and $T\approx 300$K, 
and using the same assumption on the interaction range and strength we have $\bar{A} =10^{-1} $ and:
\begin{align*}
s_\zeta&=\mathcal{O}(1\, \mathrm{nm})\\
s_x &=\mathcal{O}(1\text{ to } 10 \mathrm{nm})\\
s_t &=\mathcal{O}(10^{-6} \mathrm{s})\\ 
s_P &= \mathcal{O}(\mathrm{M\,Pa})\\ 
s_\eta &=\mathcal{O}(10^{-2} \mathrm{Pa\, s}) \, .
\end{align*}

As an illustrative example for the use of these scaling factors,
simulations were performed in a box of normalized width $100$ 
with an initial distance equal to $1$ between the dissolving crystal and the substrate. 
For both cases of calcite and sodium chlorate, this corresponds to 
thicknesses of the order of the nanometer. In addition, contact 
widths are $\sim 1\mu\mathrm{m}$ for  calcite, and $\sim 100$nm to $1\mu\mathrm{m}$
for sodium chlorate.

Some remarks are in order. 
First, the order of magnitude of the relevant pressures does not
depend much on the system.
In contrast, the order of magnitude of the timescale and 
of the relevant viscosities depend strongly on the solubility $c_0$, 
which can vary by many orders of magnitude from one material to another.

As discussed previously for dissolution with singular (power-law) repulsions, 
one could discriminate between diffusive and hydrodynamic regimes. 
The simulation results show that the high viscosity regime (hydrodynamic regime) 
is expected for  $\bar{\eta}\geq 1$  (top panel of \cref{fig:vel_force_powerLaw_2d}) 
for  $F_{Cz}/S\sim 10^2\mathrm{MPa}$ to $10\mathrm{GPa}$ with $S = \pi R^2$, and
 micro-metric crystals ($\bar{R} =100 \leftrightarrow R = 1\mathrm{\mu m}$). 
For calcite this would be expected for $\eta\sim 10^2\,\mathrm{Pa\, s}$ which is much larger of the value for water ($\approx \mathrm{mPa\,s}$). As a consequence for this system the observation of such regime should be difficult in natural environments. 
However, for highly soluble salts such as NaClO$_3$, we would need $\eta\sim 10\mathrm{mPa}$ 
much closer to the value of water. 
Therefore hydrodynamic dissolution regime should be easier to observe in this type of systems.

However, physical parameters such as viscosity and diffusion can also depend on pressure, temperature, pH or be affected by phenomena inherent to confinement. For example, large pressures are know to lead to variation of the viscosity~\cite{Wonham1967} while nano confinement when double layer is present on the surfaces, could promote higher effective viscosities (electroviscosity) \cite{Wang2010}.

One should keep in mind that there are limits in the application of our continuum model.
For instance, when $\zeta_0$ reaches the molecular scale, the continuum approach will break down and one should resort to different models based on molecular methods. An interesting step in this direction was recently proposed in the literature using Kinetic Monte Carlo simulations \cite{Hogberget2016}.
Atomistic simulations may also allow one to
tackle discontinuities of the surface profile
such as atomic steps, which where shown to be 
relevant for pressure solution experiments~\cite{Pachon-Rodriguez2011}. 

Moreover, one of the approximations used in our 
study of pressure solution is the linearization of the Gibbs-Thomson relation. 
The full nonlinear expression of the Gibbs-Thomson relation must be kept when $U'(\zeta)\ll k_BT/\Omega$.
At room temperature $k_BT/\Omega\sim 1$MPa for molecular crystals, and $k_BT/\Omega\sim 1$GPa for atomic crystals. 
As discussed at the beginning of this section, we may assume
maximum disjoining pressures $U'$ of the order of the MPa, and
the assumption $U'(\zeta)\ll k_BT/\Omega$ although not systematically valid,
should apply in many cases.
As discussed in~\cref{appendix:nonlin_GibbsThomson}, 
our analysis can be extended to the case  where the full 
nonlinearity of the Gibbs-Thomson relation is kept. 
This leads to similar results as those discussed above
in the presence of an exponential potential. 
The only  important difference appears for power-law interactions, the 
functional form of the dissolution rate 
and minimum distance with the force are not power-law anymore.
Instead, they exhibit an essential singularity as discussed
in~\cref{appendix:nonlin_GibbsThomson}.

Another limitation of our model
is the absence of elastic or plastic displacements in the solid.
However, our results show that even in the absence of elasticity or plasticity,
significant shape changes can be observed in contact zones
due to dissolution or growth kinetics in the presence of
disjoining pressure effects. 
Hence, elasticity or plasticity are not the only pathways 
towards flat contact shapes in pressure solution, and
dissolution alone is a sufficient mechanism. 
Beyond displacements, elasticity also gives rise to 
an additional contribution to the chemical potential~\cite{Pimpinelli1998}
$\sim \Omega \sigma^2/2E$, where $E$ is the Young modulus.
For this contribution to be dominant
as compared to that coming from disjoining pressure $\Omega U'$, 
one should have stresses larger than $(2EU')^{1/2}$.
Taking $U'\sim$MPa, and $E\sim 10$GPa, 
we obtain that stresses should typically exceed $10^2$MPa
for elastic effects to be relevant in the chemical potential.
In addition, pointy morphologies such as those 
obtained in our model for finite repulsions 
should lead to a concentration of stresses
which could result in significant elastic or plastic
effects. Further studies in this direction
are needed. 

Finally, one major assumption of our study is
the constant size of the contact region. While specific
needle-like crystal shapes may indeed present
a constant contact area during dissolution, it is clear
that more general shapes, e.g. conical or spherical crystals would exhibit a
growing contact area as dissolution proceeds. 
In addition redeposition of material ouside
the contact could also change the contact area during pressure solution.
Our description could still hold
if the change in the contact area was slower than
the relaxation of the crystal profile within the contact.
Such a separation of timescales, where a steady-state is reached within 
the contact as if the contact size was
constant at all times, will be denoted as the quasistatic approximation.

In the following, we discuss the validity of this approximation.
Effects such as redeposition, growth, or dissolution
outside the contact are assumed to be smaller than
the dissolution in the contact region.
From dimensional analysis of \cref{eq:height_2d}
 neglecting the contribution of surface tension,
the relaxation time $t_{relax}$ 
towards a stead-state profile $\zeta_s(r)$ 
with a contact of size $R_c$
is $t_{relax}\sim R_c^2/(D_e \tilde U'(\zeta_s))$.
In addition from force balance \cref{eq:vel_2d}, we have
$F\sim R_c^2 U'(\zeta_s)$. 
Since $U'(\zeta_s)\sim \tilde U'(\zeta_s)$,
we find $t_{relax}\sim R_c^4/(D_e F)$.
Assuming a small contact angle $\theta_{ext}$ at the edge of the contact,
dissolution induces a growth velocity for the contact radius
$dR_c/dt=u_{Cz}/\theta_{ext}$.
We must therefore require that the relaxation time is
smaller than the time associated with the growth
of the contact radius: $t_{relax}\ll R_c/(dR_c/dt)$,
leading to $ R_c^4/(D_e F)\ll R_c\theta_{ext}/u_{Cz}$.
For example in the case of a power-law potential in the diffusion-dominated regime, 
$u_{Cz}$ is given by \cref{eq:diff_2d}, and this condition leads to 
$F\gg A/(\theta_{ext}^{n+1}R_c^{n-1})$. Using the 
relation stated above in this subsection $A\sim U'\lambda^{n+1}$,
and the force balance $F\sim R_c^2 U'$,
we finally obtain a simple condition
$\lambda/R_c \ll\theta_{ext}$. Since we assumed
$\lambda/R_c\sim 10^{-5}$ above (with
$\lambda\sim$nm and $R_c\approx 100\mathrm{\mu m}$), this result suggests that
for contact angles not too small $\theta_{ext}\gg 10^{-5}$,
the quasistatic approximation should be valid. 

Within this approximation, the dissolution rate
will depend on the shape of the dissolving solid.
For example for a cone of half angle $\theta_{cone}$,
assuming no redeposition outside the contact region,
the radius of the contact area obeys $dR_c/dt=u_{Cz}\tan\theta_{cone}$. 
Similarly, for a sphere of radius $R_0$, we have
$dR_c/dt=u_{Cz}(R_0^2/R_c^2-1)^{1/2}$.
Since $u_{Cz}\sim R_c^{\beta_u}$ from~\cref{eq:summary_2d},
we find that $R_c\sim t^{1/(1-\beta_u)}$ and $u_{Cz}\sim t^{\beta_u/(1-\beta_u)}$ 
at constant force in the conical case,
and $R_c\sim t^{1/(2-\beta_u)}$ and $u_{Cz}\sim t^{\beta_u/(2-\beta_u)}$
at constant force in the spherical case when $R_c\ll R_0$.
Choosing again the example of power-law repulsion
in the diffusion limited regime where $\beta_u=-(4n+2)/(n+1)$,
we find $u_{Cz}\sim t^{-(4n+2)/(5n+3)}$
and $u_{Cz}\sim t^{-(2n+1)/(3n+2)}$
for the conical and spherical cases 
respectively.

\subsection{Comparison with existing models and experiments}

Since it relates deformation strains, contact size 
and stress on single contacts dissolution 
(eventually connecting it to the overall grain compaction problem) 
in an axisymmetric geometry, Weyl's model \cite{Weyl1959}
is a first natural candidate for comparison to our model. Weyl predicts that $u_{Cz}= 8D\lambda bF_{Cz}/R_c^2$ where D is the diffusion constant, $\lambda$ is the film thickness, $b$ a linear stress coefficient linking local solute concentration with the applied stress and $R_c$ is the contact size. 

Other models consider the phenomena at the scale of the grain 
rather than the contact region \cite{Coble1963,Elliott1973,Raj1982,Kruzhanov1998,Fowler1999}. 
Rutter~\cite{Rutter1976} summarizes most of the previously 
cited models (for diffusion controlled kinetics) 
and also treats the global problem at the thin film contact area, 
as done by Weyl. 
In cylindrical symmetry and for small external stresses,
Rutter~\cite{Rutter1976} predicts $u_{Cz}=32C_0D wVF_{Cz}/(\mathcal{R}_gT\rho_C d^3)$, while for high external stresses ($>100\,\mathrm{MPa}$) Rutter finds $u_{Cz}= 40c_0Dw\exp[F_{Cz}V/(2.3\mathcal{R}_gT)]/(d^3\rho_C)$ where $c_0$ is the concentration at the interface, $\rho_C$ is the crystal density, $D$ is the diffusion at the grain boundary, $w$ is an effective width, $\mathcal{R}_g$ is the gas constant and $d$ is the grain size (proportional to the contact size).  

The relations predicted by Weyl and Rutter are in general 
not in agreement with our predictions both for power law repulsion 
and finite exponential repulsion \cref{eq:hydro_2d,eq:diff_2d,eq:vel_exp_2d}.

Globally, the absence of description of microscopic physical ingredients such as viscosity, 
interaction potential, and surface tension in these models lead to a very different and non-specific behavior.

Previous modeling attempts have also
addressed  the regime of slow interface kinetics~\cite{Kruzhanov1998}. They suggest
that the dissolution rate could then be independent
of the contact area.
The investigation of this limit is
an interesting perspective for further development of our model.

A number of experimental observations have suggested power law 
relations between strain rates (crystal velocity) 
and applied stress and or grain size\cite{Zhang2005,Gratier2013,DenBrok1999}. 
This is compatible with the results we obtained for the singular repulsive power-law potential in \cref{eq:hydro_1d,eq:hydro_2d,eq:diff_1d,eq:diff_2d,fig:vel_force_powerLaw_2d}.
However, \citet{Croize2010a}~underline that though 
there exists a positive correlation between 
the strain rate and the applied stress, 
this dependence is weak. With the support 
of both original measurements on calcite 
pressure solution and data from the literature, 
they claim that other effects such as the grain size are likely to be dominant.
These observations are consistent with the scenario predicted for exponential interaction in \cref{eq:vel_exp,fig:exp_vel}.

Using the pressure range $1\text{ to } 10^3\,\mathrm{MPa}$, which is the one usually considered in pressure solution experiments,
the velocities (dissolution rates) 
obtained by our simulations are $10^{-3}\text{ to } 10^{-1} \mathrm{nm\,s^{-1}}$ for calcite 
and $10^{-1}\text{ to } 10\, \mathrm{\mu m \, s^{-1}}$ for sodium chlorate. 
The observable usually reported in pressure solution experiments is the strain rate.
Experimental values of the strain rates for calcite \cite{Croize2010a,Zhang2005} 
vary between $10^{-9}\mathrm{s}^{-1}$ and $10^{-4}\mathrm{s}^{-1}$. 
Using  $\dot{\epsilon} = u_{Cz}/R_c$ as the definition of the strain rate \cite{Croize2010a}, 
we obtain 
values between $10^{-6}\mathrm{s}^{-1}$ and $10^{-4}\mathrm{s}^{-1}$, 
compatible with the experimental ones.
For NaClO$_3$, because of the faster time scales 
due to the much higher solubility, the dissolution rate 
and as a consequence the strain rate increases of a factor of about $10^5$. 
This is in disagreement with the literature \cite{DenBrok1999}, 
where similar orders of magnitude as those of calcite are found. 
Such discrepancy 
could be caused by the fact that in our system 
exhibits an under-saturated concentration bath at the boundaries
of the contact.
In multi-contact systems where the liquid reservoir
per contact is finite, the global
supersaturation of the bath should increase due to 
the release of crystal molecules in the liquid.
This should lead to a decrease of the dissolution rates.
The study of such interactions between different contacts 
is therefore an important perspective for our modeling approach
to address systems with multiple contacts.

As far as the morphology of the contact is concerned, 
some experiments on quartz grains aggregates \cite{Cox1991} 
showed that in addition to relative smooth interfaces, 
irregular ridge and plateau structures can develop at the grain contacts
 after undergoing pressure solution. The appearance of point-like and ridge-like
singularities for exponential repulsions in our model could be 
a first step towards the understanding of these morphologies.

In general, further experimental investigation  involving observations at the scale 
of one microscopic contact would be useful to test our model predictions.

\section{Conclusions}

In conclusion, we have presented a thin film model
for the dynamics of lubricated contacts during dissolution
and growth under load, accounting for surface tension, interactions, diffusion, and 
hydrodynamics.  This model describes the coupled evolution
of the space-dependent pressure field  $p$ in the liquid, and 
the film thickness $\zeta$ via
\cref{e:2D_mass_cons_hydro_p,e:2D_mass_cons_c_lubric_fastkin_p}. 
An additional constraint originating in global 
force balance \cref{eq:force_balance_lubri,eq:force_balance_lubri_xy}
determines the crystal velocity $\mathbf u_C$.

Using this model, we have discussed the dynamics of pressure
solution for single contacts 
of fixed or slowly varying size and
with symmetric geometries,
using some simplifying assumptions.
We find that the dissolution rate and  contact morphology
exhibit distinctive behaviors depending
on the finiteness of the repulsion at contact.
Furthermore, we find that crystal-substrate touching contact 
is never reached in steady-state
for any load when viscosity and surface tension
are taken into account.

Much yet remains to be done to explore
the different regimes emerging for our thin film model. However, this model paves
the way for a systematic and physically consistent analysis
of the influence of different microscopic ingredients
on pressure solution and growth in confined environments.

\section*{Acknowledgements}

The authors wish to thank Alois Meckenstock for 
useful comments on the manuscript.
This project has received funding from the European Union's Horizon 2020 research and innovation program under grant agreement No \textbf{642976}.

\appendix


\section{Identities resulting from translational invariance of the free energy}
\label{appendix:force_balance}

Here we derive some integral identities
 that are used in the main text. 
These identities express the fact that
the total force resulting from a translational invariant energy must vanish.

Consider a generic free energy functional $\mathcal{F_D}$ 
acting over a domain $\mathcal{D}$ in $d$ dimensions
 and with boundary $\partial\mathcal{D}$ in $(d-1)$ dimensions.
Let us assume that its variation can be written as a surface integral
\begin{equation}
\delta \mathcal{F} = \int_{\partial\mathcal{D}} \mathrm{d}S\, (\delta \mathbf{r}\cdot \hat{\mathbf{n}})\frac{\delta \mathcal{F_D}}{\delta\mathbf{r}} \, ,
\end{equation}
where $\delta\mathbf{r}$ is a $d$-dimensional infinitesimal variation of the domain boundary. 

Assume now  that $\mathcal{F_D}$ 
is invariant under translations. 
Then, $\delta \mathcal{F}$ must vanish under infinitesimal translations, 
{\it i.e.} when $\delta \mathbf{r}=\mathrm{d}\mathbf r$ is an arbitrary constant (independent on space coordinates).
As a consequence
\begin{equation}
0 =\mathrm{d}\mathbf r \cdot \int_{\partial\mathcal{D}} \mathrm{d}S\, \hat{\mathbf{n}}\, \frac{\delta\mathcal{F_D}}{\delta\mathbf{r}}\, .
\end{equation}
Since this is true for any $\mathrm{d}\mathbf r$, we find that
the force acting on the domain surface vanishes:
\begin{equation}
0 = \int_{\partial\mathcal{D}} \mathrm{d}S\, \hat{\mathbf{n}}\, \frac{\delta\mathcal{F_D}}{\delta\mathbf{r}}\, .
\end{equation}
This relation is 
valid for arbitrary shapes of the domain $\mathcal{D}$.

In particular, consider the surface energy
\begin{equation}
\mathcal{F_S} = \int_\mathcal{\partial\mathcal{D}} \mathrm{d}S\, \gamma(\hat {\mathbf n} )\, ,
\end{equation}
whose variation is given by
\begin{equation}
 \int_\mathcal{\partial \mathcal{D}} \mathrm{d}S \, \hat{\mathbf{n}}\,
(\kappa:\tilde{\gamma}) = 0 \, ,
\end{equation}
where $\gamma$ is a general surface tension (function of the orientation), $\tilde{\gamma}$ is the stiffness tensor and $\kappa$ is the curvature tensor.
In the special case where the surface tension
is isotropic, i.e.  $\gamma$ does not depend on $\hat{\mathbf n}$,
we obtain a known equality: the integral of the 
mean curvature times the normal vector of an arbitrary
(sufficiently regular) surface vanishes\cite{Blackmore1985}
\begin{equation}
 \int_\mathcal{\partial \mathcal{D}} \mathrm{d}S \, \hat{\mathbf{n}}\;H = 0 \, ,
\end{equation}
where $H$ is the mean curvature.

Finally another useful relation is obtained when choosing 
an energy proportional to the volume of the domain $\mathcal{ D}$:
\begin{equation}
\label{eq:boundary_integral_normal}
\int_\mathcal{\partial \mathcal{D}} \mathrm{d}S \, \hat{\mathbf{n}} = 0\, .
\end{equation}
We find that the integral of the normal vector vanishes
on any closed regular surface.

\section{Rescaling and units}
\label{appendix:scaling}

Let us recall the type of substrate-crystal interactions considered, \cref{eq:interaction_exp,eq:singular_pot}:
\begin{align*}
U(\zeta) &= \frac{A}{\zeta^n} \quad\text{Singular at contact},\\
U(\zeta) &= Ae^{-\frac{\zeta}{\lambda}}\quad\text{Finite at contact} .
\end{align*}
For simplicity we only show the scaled equations in 1D.
In the case of the power law repulsion, \cref{eq:singular_pot} with $n=3$, we have
\begin{subequations}
\label{eq:height_scaled_pl}
\begin{align}
&\partial_{\bar{t}} \bar{\zeta} = -\partial_{\bar{x}}\Bigl[ \bar{\zeta} \partial_{\bar{x}} (\partial_{\bar{x}\bar{x}} \bar{\zeta}+\frac{1}{\bar{\zeta}^4})\Bigr] - \bar{u}_{Cz} \, ,
\\
&\bar{u}_{Cz} \, \int_0^{\bar{L}} \mathrm{d}\bar{x}\, \int_{\bar{x}}^{\bar{L}} \mathrm{d}\bar{x}'\, \frac{24\bar{\eta} \bar{x}'}{\bar{\zeta}^3} = \bar{F}_{Cz} + 2\int_0^{\bar{L}} \mathrm{d}\bar{x}\, \frac{1}{\bar{\zeta}^4} \, ,
\end{align}
\end{subequations} 
where $\bar{u}_{Cz}$, $\bar{\eta}$ and $\bar{F}$ are the
rescaled velocity, viscosity and external force, respectively.
For the exponential repulsion \cref{eq:interaction_exp}, we have
\begin{subequations}
\label{eq:height_scaled_exp}
\begin{align}
&\partial_{\bar{t}} \bar{\zeta}=- \partial_{\bar{x}}\Bigl[ \bar{\zeta} \partial_{\bar{x}} (\partial_{\bar{x}\bar{x}} \bar{\zeta} +e^{-\bar{\zeta}})\Bigr] - \bar{u}_{Cz}\, ,\\
&\bar{u}_{Cz} \, \int_0^{\bar{L}} \mathrm{d}\bar{x}\, \int_{\bar{x}}^{\bar{L}} \mathrm{d}\bar{x}'\, \frac{{{24}}\bar{\eta} \bar{x}'}{\bar{\zeta}^3} = \bar{F}_{Cz} + 2\int_0^{\bar{L}} \mathrm{d}\bar{x}\, e^{-\bar{\zeta}} \, .
\end{align}
\end{subequations}

If $y$ is an arbitrary variable and $\bar{y}$ its 
normalized counterpart used in simulations, we define $s_y$ the scaling factor 
that has to be applied 
to recover the natural variables from the normalized
simulation variables: $y=s_y \bar{y}$.
Defining $\bar{A}$ as a non dimensional quantity equal to $A/\gamma$ for the exponential repulsion, and equal to $A/(\gamma\lambda^3)$ for the power-law repulsion, the scaling factors are for both \cref{eq:height_scaled_pl,eq:height_scaled_exp}:
\begin{subequations}
\label{eq:scalings}
\begin{align}
s_\zeta &=\lambda \\
s_x &=\lambda(\frac{1}{\bar{A}})^{1/2}\\
s_t &=\lambda^3 \frac{k_BT}{D\Omega^2c_0\gamma\bar{A}^2}\\
s_F^{1D} &=\gamma \bar{A}^{1/2}  \\
s_F^{2D} &= \gamma\lambda  \\
s_p &= \frac{\gamma \bar{A}}{\lambda}\label{eq:scaling_pressure}\\
s_\eta &= \lambda^2\frac{k_BT}{D\Omega^2c_0} \label{eq:scaling_visc}\\
s_{u_{Cz}} & = \frac{D\Omega^2c_0\gamma\bar{A}^2}{k_BT\lambda^2}\, .
\end{align}
\end{subequations}
The superscripts $1D$ and $2D$ explicitly indicate those 
scalings which differ in the ridge and axisymmetric system.
Also note that
$s_p = s_F^{1D}/ s_x $ in 1D, 
while $s_p = s_F^{2D}/s_x^2$ in 2D.

\section{Steady state in the absence of surface tension}
\label{appendix:derivation_noSurfTension}

We here illustrate how to derive some relations of \cref{sec:power_law,sec:exp_noSurf} 
for the axisymmetric system (2D). An analogous procedure can be followed in 1D.
In 2D, a single integration of \cref{eq:steady_state} leads to 
\begin{equation}
\label{eq:single_int}
\frac{r}{2}u_{Cz} = D_{e}\zeta\partial_r\zeta U''(\zeta) \, .
\end{equation}
Using the previous relation to express the differential $r\mathrm{d}r$ as a function of $d\zeta$ 
and considering only the contribution of the contact area, we can rewrite \cref{eq:vel_2d} in a more convenient form:
\begin{multline}
\label{eq:force}
F_{Cz}=2\pi\int_0^{R_c} r 12\eta D_{e} \,\mathrm{d}r \int_{\zeta(r)}^{\zeta(R_c)}\mathrm{d}\zeta\, \frac{U''(\zeta)}{\zeta^2}\\
-2\pi\int_0^{R_c}\mathrm{d}r\, r U'(\zeta(r))\, . 
\end{multline}

\subsection{Singular power law repulsion}
\label{appendix:noSurf_pwLaw}

\begin{figure}
\includegraphics[width=\linewidth]{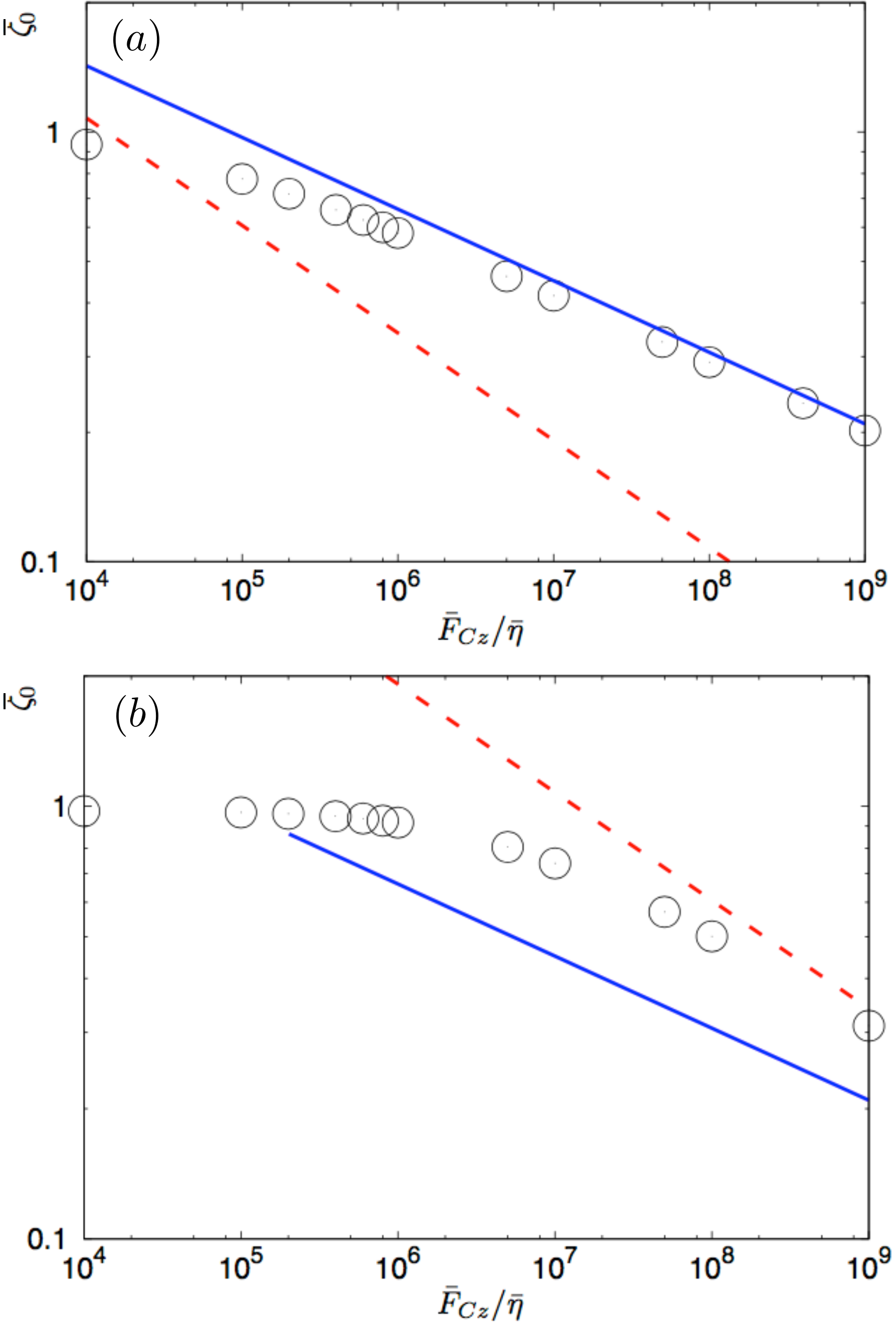}
\caption{Minimum film thickness $\zeta_0$ 
as a function of the applied load for (singular) power law repulsion. 
Geometry: axisymmetric contact. 
Lines represent analytical predictions extracted from \cref{eq:F_vs_zita_singularRep:2d} with $\bar{R}_c = \bar{R} =100$, 
circles indicate simulation results. Solid blue line, hydrodynamic regime;  dashed red line, diffusion regime. 
(a) $\bar{\eta} =1$; 
(b) $\bar{\eta} = 10^{-3}$. 
The results are given in normalized units.
}
\label{fig:zita_force_powerLaw2d}
\end{figure}
Using \cref{eq:profile_powerLaw} together with \cref{eq:singular_pot} in \cref{eq:force}, we find
\begin{equation}
\label{eq:force_hydro+diff}
\begin{split}
F^{2D}_{Cz} &= 12\eta C_1(r_m,R_c)\, (R_c^2u_{Cz}) ^{\frac{n+3}{n}}+ \\
& \qquad C_2(r_m,R_c) \,(R_c^2u_{Cz}) ^{\frac{n+1}{n}} \, ,
\end{split}
\end{equation}
where 
\begin{align*}
C_1&=D_{e}\frac{n(n+1)\pi A}{n+3}\Bigl\{\frac{-R_c^2}{\zeta^{n+3}(R_c)} +\\
& \frac{n r_m^2}{(2n+3)\zeta_0^{n+3}}\Bigl[(\frac{R^2_c}{r_m^2} -1)^{\frac{n+3}{n}}+ 1\Bigr ] \Bigl ( \frac{r^2_m/R_c^2}{4D_{e}(n+1)A}\Bigr)^{\frac{n+3}{n}}\Bigr\}\\
C_2&=\frac{\pi n^2 A r^2_m}{(2n+1)}\Bigl [(\frac{R_c^2}{r_m^2} -1)^{\frac{2n+1}{n}} +1 \Bigr]\Bigl ( \frac{r^2_m/R_c^2}{4D_{e}(n+1)A}\Bigr)^{\frac{n+1}{n}} \, .
\end{align*}
For large external loads, we have $r_m\sim R_c$
 and $\zeta(r = R_c)\gg \zeta_0$, leading to \cref{eq:F_vs_v_singularRep:2d}.

The constants used in the main text in \cref{eq:hydro_1d,eq:hydro_2d,eq:diff_1d,eq:diff_2d} 
were obtained considering that one of the two terms in \cref{eq:force_hydro+diff} dominates
in the force balance depending on the value of the viscosity. Their 
expressions are
\begin{align}
C^{1D}_{h} &=\frac{2\Bigl(D_eA(n+1)\Bigr)^{\frac{3}{n+3}}}{\Bigl(\frac{24n\sqrt{\pi}}{n+3}\phi(\frac{n+3}{n})\Bigr)^{\frac{n}{n+3}}} \\
C^{2D}_{h} &= \frac{4\Bigl(D_eA(n+1)\Bigr)^{\frac{3}{n+3}}}{\Bigl( \frac{12\pi n^2}{(2n+3)(n+3)}\Bigr)^{\frac{n}{n+3}}}\\
C^{1D}_{d} &=\frac{2D_eA^{\frac{1}{n+1}}(n+1)}{\Bigl(2n\sqrt{\pi}\phi(\frac{n+1}{n}) \Bigr)^{\frac{n}{n+1}}} \\
C^{2D}_{d} &= \frac{4D_eA^{\frac{1}{n+1}}(n+1)}{\Bigl(\frac{\pi n^2}{2n+1}\Bigr)^{\frac{n}{n+1}}}\, .
\end{align}
As discussed in the main text, the force can also be 
written as a function of the distance $\zeta_0$ 
between the substrate and  the crystal surface at the center of the contact:
\begin{multline}
\label{eq:F_vs_zita_singularRep:2d}
\frac{F^{2D}_{Cz}}{\pi R_c^2} =12\eta D_{e}\frac{n^2(n+1) A}{(2n+3)(n+3)}\Bigl (\frac{1}{\zeta_0}\Bigr )^{n+3}\\
 + \frac{n^2 A}{(2n+1)}\Bigl (\frac{1}{\zeta_0}\Bigr )^{n+1} \, ,
\end{multline}
leading to the asymptotic scaling reported in \cref{sec:summary}.
These results are confirmed by the numerical solution as showed in \cref{fig:zita_force_powerLaw2d}.

\begin{figure}
\includegraphics[width=\linewidth]{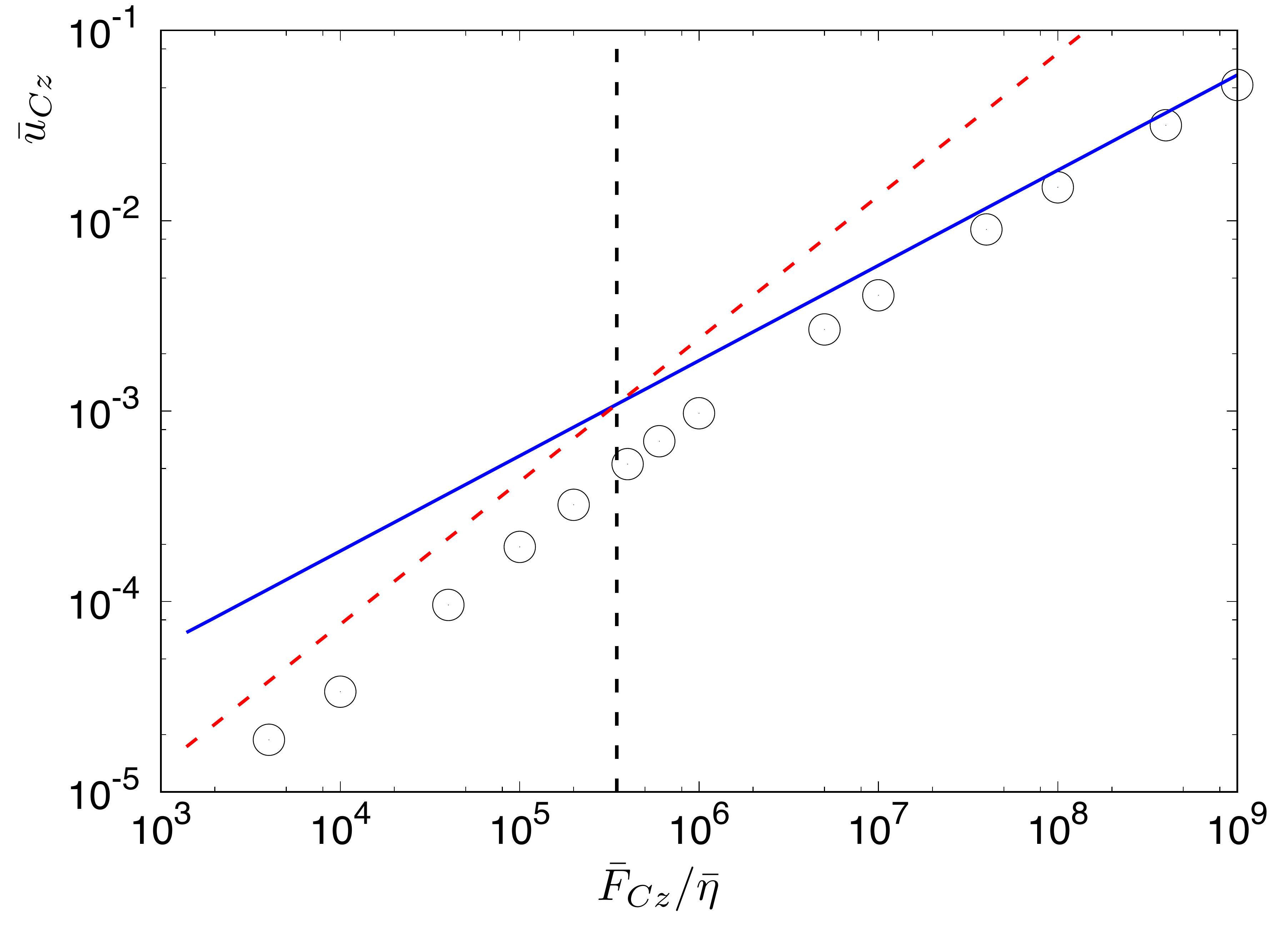}
\caption{Dissolution rate as a function of the applied load for power law (singular) repulsion. 
Geometry: axisymmetric contact. The viscosity is $\bar{\eta} = 10^{-1}$.  
Lines represent analytical predictions with $\bar{R}_c = \bar{R} =100$, circles indicate simulation results. 
Solid blue line, hydrodynamic regime \cref{eq:hydro_2d};  
dashed red line, diffusion regime \cref{eq:diff_2d}. 
The black dashed line represents the expected threshold between the two regimes 
according to \cref{eq:critical_L_2d}. 
The results are shown in normalized units.
}
\label{fig:critical_size}
\end{figure}

Finally, as showed in \cref{fig:critical_size}, we have explored the transition between the diffusion and hydrodynamic scaling laws. 
This was done using an intermediate viscosity, $\bar{\eta} = 0.1$, and looking at the dissolution rates in a 2D contact of size $\bar{R}_c\approx\bar{R} =100$.
The constants appearing in \cref{eq:critical_L_1d,eq:critical_L_2d} are:
\begin{align}
B_{1D} &= \Bigl [\frac{24D_e(n+1)\phi\Bigl(\frac{n+3}{n}\Bigr)}{(n+3)\Bigl[2\phi\Bigl(\frac{n+1}{n}\Bigr)\Bigr]^{\frac{n+3}{n+1}}}\Bigr ]^{\frac{n+1}{2}} \frac{1}{nA\sqrt{\pi}}\, \\
B_{2D} &= \Bigl [ \frac{12 D_e (n+1)(2n+1)^{\frac{n+3}{n+1}}}{(2n+3)(n+3)}\Bigr ]^{\frac{n+1}{4}} \frac{1}{n\sqrt{A\pi}}\, .
\end{align}
From \cref{eq:critical_L_2d}, with $\bar{R}_* =100$, $\bar{\eta} = 0.1$, $n=3$ 
(since in simulations units $D_e =1$ and $A =1/3$, $B_{2D} \approx 5.4$) 
we expect the diffusion limited regime approximately for $\bar{F}_{Cz}<3.5\,10^4$
and the hydrodynamic regime otherwise. 
The threshold indicated in the figure by the dashed vertical line
is compatible with the observed trend.

\subsection{Finite exponential repulsion}
\label{appendix:noSurf_exp}

In the case of a finite exponential repulsion,
 manipulations similar to those presented in the previous section
lead to the following form of the force balance relation
\begin{equation}
\label{eq:force:exp:general}
F^{2D}_{Cz} = 48\eta\frac{\pi D_{e}^2A^2}{\lambda^3u_{Cz}}\psi(\frac{\zeta_0}{\lambda}) + \frac{\pi D_{e}A^2}{\lambda}\Bigl (\frac{2\zeta_0}{\lambda}+1 \Bigr)\frac{e^{-\frac{2\zeta_0}{\lambda}}}{u_{Cz}}\, ,
\end{equation}
with
\begin{equation}
\label{eq:psi}
\psi(z_0) =\lambda\int_{z_0}^\infty \mathrm{d} z\, e^{-z} 
\Bigl (e^{-z} + zEi(-z)\Bigr) \, ,
\end{equation}
where $E_i$ is the exponential integral defined as
\begin{equation}
Ei(x) = -\int_{-x}^\infty \frac{e^{-s}}{s} \mathrm{d}s \, .
\end{equation}
Inserting the expression  of $u_{Cz}$ from \cref{eq:vel_exp_2d} 
into \cref{eq:force:exp:general} we obtain \cref{eq:force_exp2d}.

\subsection{1D case}

In 1D, the derivations are similar to the 2D case.
We obtain \cref{eq:force_exp1d} where the two
functions $\tilde{\psi}_1$ and $\tilde{\psi}_2$ are defined as:
\begin{align}
\label{eq:psi_1}
\tilde{\psi}_1(z_0) &=\int_{z_0}^\infty\frac{e^{-z}(e^{-z} + z E_i(-z))}{[(1+z_0)e^{-z_0} - (1+z)e^{-z}]^{\frac{1}{2}}}\mathrm{d}z\, , \\
\label{eq:psi_2}
\tilde{\psi}_2(z_0)&=\int_{z_0}^\infty\frac{ze^{-2z}}{[(1+z_0)e^{-z_0} - (1+z)e^{-z}]^{\frac{1}{2}}}\mathrm{d}z \, .
\end{align}

\section{Surface tension contribution in finite repulsion}
\label{appendix:derivation_expTip}

\begin{figure}
\includegraphics[width=\linewidth]{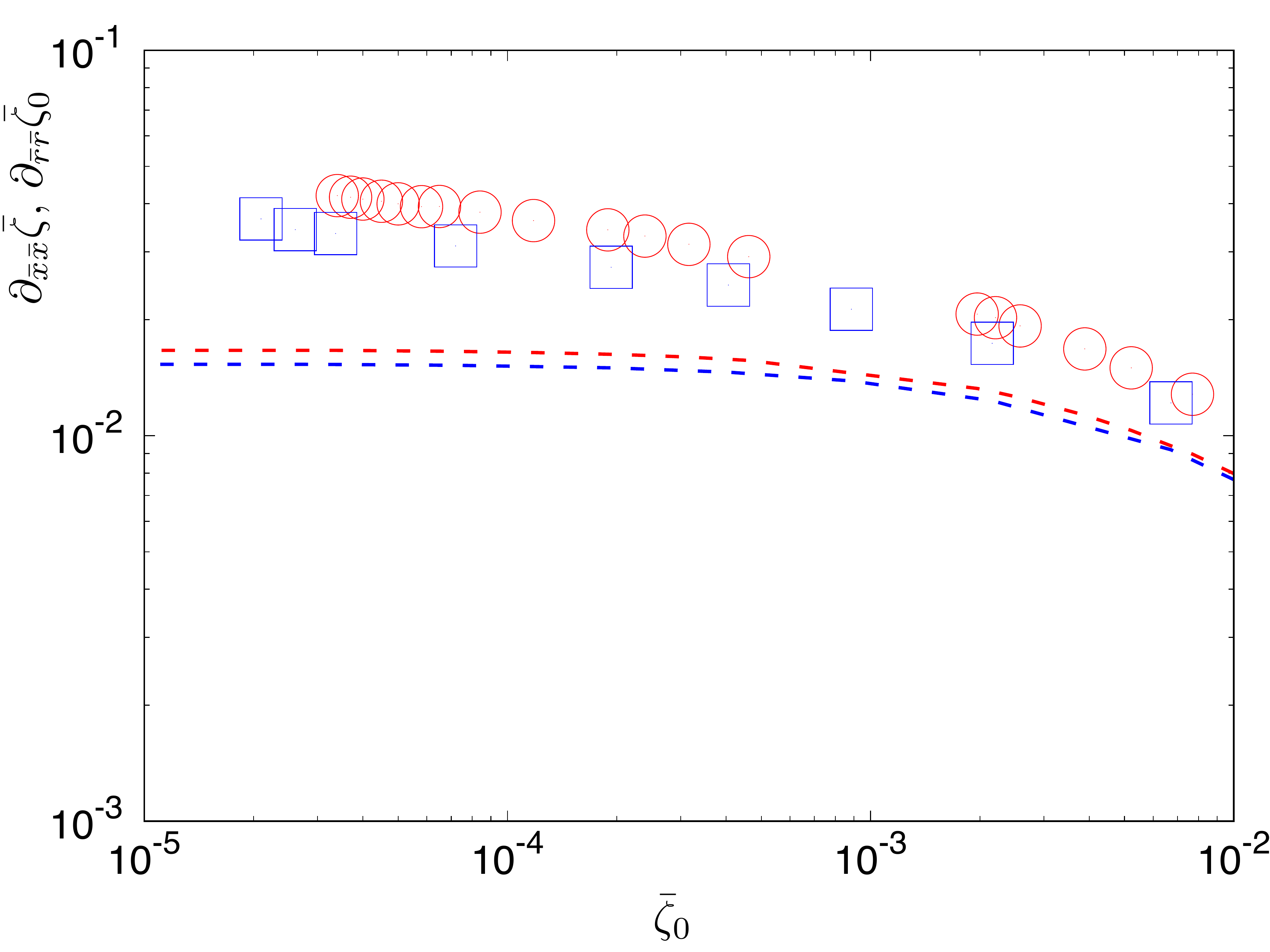}
\caption{Curvature at the tip as a function of the tip-substrate gap $\bar{\zeta}_0$.
Red circles: simulations result for the wedge contact (1D); 
Blue squares: simulations result for the axisymmetric conical contact (2D). 
Sizes of the simulation boxes are $\bar{L}=\bar{R}=100$; 
dashed red and blue lines analytical prediction using the solution of \cref{eq:interpolation_2d},\cref{eq:interpolation_1d}, respectively using the the assumption $L_c\approx L$, $R_c \approx R$. 
The results are in normalized units.}
\label{fig:curv_vs_zita}
\end{figure}

We here report a derivation of the relation between the second derivative 
of the interface $\partial_{xx}\zeta_0$ or $\partial_{rr}\zeta_0$ and the minimum film width $\zeta_0$.
This relation is obtained through a procedure where we match 
the two approximate solutions at the tip
$\zeta_{tip}$ in \cref{eq:tip_parabola}, 
and far from the tip $\zeta_{sing}$ in \cref{eq:pointy_shape}.

\subsection{1D case}
In 1D we proceed as follows.
Integrating two times \cref{eq:steady_1d} we have
\begin{equation}
\label{eq:steady_state_1d_integrated}
0=\frac{x^2}{2D_e}u_{Cz} - \frac{A}{2\lambda}(\zeta^2-\zeta_0^2)+
\gamma \Bigl(\zeta\partial_{xx}\zeta-\zeta_0\partial_{xx}\zeta_0  - \frac{1}{2}(\partial_x{\zeta})^2\Bigr)\, ,
\end{equation}
where we used the parity condition $\partial_x\zeta_0 = 0$ and the expansion of $\tilde{U}$,
(given for the exponential repulsion by \cref{eq:exp_rep}), 
up to second order in $\zeta$: $\tilde{U} \approx  A(-1 + \zeta^2/(2\lambda^2))$.

Adding a perturbation $\delta\zeta$ to $\zeta_{sing}=\omega |x|$ given by \cref{eq:pointy_shape_1d}
we have
\[
\zeta_{far}=\zeta_{sing} + \delta\zeta\, ,
\]
with 
\[
\omega = \Bigl (\frac{u_{Cz}}{D_eA}\Bigr )^{\frac{1}{2}}\lambda\, .
\]
We then insert this relation in \cref{eq:steady_state_1d_integrated} to determine $\delta\zeta$
far from the tip.
Neglecting the terms of smaller than $\delta\zeta$ for large $x$
we find
\begin{equation}
\label{eq:delta_1d}
\delta \zeta = \gamma\frac{-\frac{1}{2}\omega^2-\zeta_0\partial_{xx}\zeta_0}{ \frac{A}{\lambda^2}\omega x}\, .
\end{equation}

We define $x_*$ as the value of $x$ at which we match
the solutions $\zeta_{sing}$ and $\zeta_{tip}$.
 We obtain two independent relations.
The first one accounts for the matching of the 
surface profiles  at $x=x_*$,
leading to $\omega x_* +\delta \zeta(x_*) = \zeta_0 + \partial_{xx}\zeta_0 x_*^2/2 $.
The second relation comes from
the matching of the slopes $\omega +\partial_x\delta \zeta(x)|_{x_*} = \partial_{xx}\zeta_0 x_*$.
Combining the two relations we obtain the following system of equations
\begin{equation}
\label{eq:interpolation_1d}
\begin{split}
\frac{3}{2}\partial_{xx}\zeta_0x_*^2-2\omega x_*+\zeta_0&=0,\\
\omega x_* -\gamma\Bigl(\frac{\omega^2}{2}\zeta_0\partial_{xx}\zeta_0\Bigr)\frac{\lambda^2}{A\omega x_*}
 &= \zeta_0 + \frac{\partial_{xx}\zeta_0}{2}x_*^2\, .
\end{split}
\end{equation}

\subsection{2D case}
In the axisymmetric system (2D) we follow a similar procedure. 
However, extra terms connected to the different expression of the curvature appear.
Following the same steps as for the derivation
of \cref{eq:steady_state_1d_integrated}, we \cref{eq:steady_2d} two times. Then, 
given the parity condition and the expansion 
of $\tilde{U}$ for small $\zeta$, we obtain
\begin{equation}
\label{eq:steady_state_2d_integrated}
\begin{split}
&0=\frac{r^2}{4D_e}u_{Cz} - \frac{A}{2\lambda}(\zeta^2-\zeta_0^2)\,+\\
&\gamma \Bigl(\zeta\partial_{rr}\zeta-2\zeta_0\partial_{rr}\zeta_0  - \frac{1}{2}(\partial_r{\zeta})^2+\frac{\zeta\partial_r\zeta}{r} - \int_{0} ^r \mathrm{d}r'\frac{(\partial_{r'} \zeta )^2}{r'}\Bigr)\, .
\end{split}
\end{equation}

To derive an explicit expression for the correction to $\zeta_{sing}$, 
we insert its expression \cref{eq:pointy_shape_2d} plus a perturbation $\delta\zeta$ in \cref{eq:steady_state_2d_integrated}.
Thus \cref{eq:steady_state_2d_integrated} becomes
\begin{equation}
\label{eq:steady_state_2d_expansion}
\begin{split}
&0= -\frac{A}{\lambda^2}\omega r \delta\zeta -\gamma\Bigl( (\partial_{rr}\zeta_0)^2\frac{r_*^2}{2} + 2\zeta_0\partial_{rr}\zeta_0\Bigr )\, + \\
&\gamma\Bigl(\frac{1}{2}\omega^2-\omega^2\ln (\frac{r}{r_*}) + \omega\frac{\delta\zeta}{r}+ 2\omega\int_{r_*} ^r \mathrm{d}r \frac{\partial_r \delta\zeta}{r} + \omega r\partial_{rr}\delta\zeta \Bigr)\, .
\end{split}
\end{equation}
where now in 2D
\[
\omega = \Bigl (\frac{u_{Cz}}{2D_eA}\Bigr )^{\frac{1}{2}}\lambda\, .
\]
Note that the integral term of \cref{eq:steady_state_2d_integrated} was rewritten as follows
\begin{align*}
\int_{0} ^r \mathrm{d}r \frac{(\partial_r \zeta)^2}{r} =
\int_{0} ^{r_*} \mathrm{d}r \frac{(\partial_r \zeta_{tip})^2}{r} +\int_{r_*} ^r \mathrm{d}r \frac{(\partial_r \zeta_{sing})^2}{r}\\
=-\gamma (\partial_{rr}\zeta_0)^2\frac{r_*^2}{2} + \omega^2\ln (\frac{r}{r_*}) + 2\nu\int_{r_*} ^r \mathrm{d}r \frac{\partial_r \delta\zeta}{r}\, .
\end{align*}

If in the matching procedure, for $r\gg r_*$, we keep only dominant terms, 
as done above for the 1D case, we would obtain $\delta\zeta\approx 1/r$. 
Since this does not behave properly we make a crude approximation 
and keep only one higher order term $\omega\delta\zeta/r$ 
to account for higher order contributions in \cref{eq:steady_state_2d_expansion}
( indeed, as $r\rightarrow r_*$ the integral term vanishes 
and we expect the other relevant term $r\partial_{rr}\delta\zeta$ 
to be of the same order as $\omega\delta\zeta/r$).
With these assumptions we find
\begin{equation}
\label{eq:delta_2d}
\delta \zeta = \gamma\frac{\frac{1}{2}\omega^2-2\zeta_0\partial_{rr}\zeta_0 -\omega^2\ln\frac{r}{r_*}- \frac{1}{2}(\partial_{rr}\zeta_0)^2r_*^2}{ \frac{A}{\lambda^2}\omega r - \gamma\omega \frac{1}{r}}\, .
\end{equation} 
As before to obtain the matching between the two solutions $\zeta_{tip}$ and $\zeta_{sing}$
we use two conditions. First,
we consider the matching of the thicknesses 
$\omega r_* +\delta \zeta(r_*) = \zeta_0 + \partial_{rr}\zeta_0 r_*^2/2$.
A second relation accounts for the matching of the slopes 
$\omega +\partial_r\delta \zeta(r)|_{r_*} = \partial_{rr}\zeta_0 r_*$. 
Since in this case $\partial_r\delta\zeta(r)$ 
does not diverge for $r\rightarrow 0$, 
and since $r_*$ is assumed to be small, 
we neglect the contribution $\partial_{r}\zeta|_{r_*}$ in the slope. 
This lead to the following system of equations:
\begin{equation}
\label{eq:interpolation_2d}
\begin{split}
&\omega r_* +\gamma\Bigl(\frac{\omega^2}{2}-2\zeta_0\partial_{rr}\zeta_0 -\frac{(\partial_{rr}\zeta_0)^2}{2}r_*^2 \Bigr )\Bigl(\frac{A}{\lambda^2}\omega r_* - \gamma\eta \frac{1}{r_*}\Bigr)^{-1}\\
 &\qquad \qquad= \zeta_0 + \partial_{rr}\zeta_0\frac{r_*^2}{2},\\ 
&\partial_{rr}\zeta_0 r_* = \omega  \, .
\end{split}
\end{equation} 

\subsection{Numerical solution}
Inserting the asymptotic analytical expression of the dissolution rate $u_{Cz}$ \cref{eq:vel_exp}, we solved the linear systems of \cref{eq:interpolation_1d,eq:interpolation_2d} using MINPACK routine\cite{More1980}.  We obtain values of $x_*$ ($r_*$) and of $\partial_{xx}\zeta_0$ ($\partial_{rr}\zeta_0$) for a given minimum distance $\zeta_0$. The results, displayed in \cref{fig:curv_vs_zita}, are represented by the dashed lines and compared with the simulation results. In particular we find (in normalized units) for $\zeta_0 = 0$, 
$\partial_{\bar{x}\bar{x}}\bar{\zeta}_0 \approx0.0167 $ 
and $\partial_{\bar{r}\bar{r}}\bar{\zeta}_0 \approx0.0153$.

\section{Beyond the linearization of the Gibbs-Thomson relation}
\label{appendix:nonlin_GibbsThomson}

A simple substitution allows one to include the effect of the 
exponential term in the analysis of the contact profile in the absence of 
surface tension:
\begin{equation}
U'(\zeta)\rightarrow  k_BT\exp[\frac{U'(\zeta)}{k_BT}].
\label{eq:nonlinear_Gibbs_Thomson}
\end{equation}
This leads to a different definition of $\tilde U$ from
the relation
\begin{eqnarray}
\tilde U'(\zeta)=\zeta U''(\zeta)\exp[\frac{U'(\zeta)}{k_BT}].
\end{eqnarray}
The same procedure as that discussed
in  \cref{sec:results}
can then be applied with this new expression for $\tilde U$. 

For power-law potentials, this leads to an essential singularity
in $\tilde U$ when $\zeta\rightarrow 0$:
\begin{equation}
\tilde U(\zeta)=
\zeta   k_BT \left(e^{-\frac{A n \zeta ^{-n-1}}{
   k_BT}}-\frac{E_{1+\frac{1}{n+1}}\left(\frac{A n \zeta ^{-n-1}}{
   k_BT}\right)}{n+1}\right)\, ,
 \end{equation}  
where 
\begin{eqnarray}
E_m(z)= \int_1^\infty \!\!\! dt \frac{{\rm e}^{-zt}}{t^m}.
\end{eqnarray}
This essential singularity
appears in the relation between $u_{Cz}$ 
and the minimum thickness $\zeta_0$ when
$\zeta_0\rightarrow 0$:
\begin{eqnarray}
u_{Cz}=4 D_e [\tilde U(\zeta_0)-\tilde U(\infty)].
\end{eqnarray}

In contrast, there is no significant
change in the case of an exponential potential.
Indeed, the central property of being finite
when $\zeta_0\rightarrow 0$ is not affected
by \cref{eq:nonlinear_Gibbs_Thomson}.
Thus, the exponential potential again leads to a pointy shape,
and constant dissolution rate obeying \cref{eq:vel_exp}.
Moreover,
the details of the regularization of the tip due
to surface tension can be affected
but we do not expect major changes.

\bibliography{biblio_pressure_sol}

\end{document}